\documentclass[review,authoryear,3p,times,10pt]{elsarticle}

\usepackage{multirow,setspace,times,amssymb,amsmath,graphicx,color,rotating,subfigure,url}
\usepackage{lineno,color}
\usepackage{natbib}
\usepackage{booktabs}%
\usepackage{longtable}%
\usepackage{rotating}
\usepackage{lscape}
\usepackage{pdflscape}
\usepackage{ulem}
\usepackage{threeparttable}
\usepackage{bm}

\graphicspath{{Figures/}}
\usepackage[table]{xcolor}
\usepackage{tabularx}
\usepackage{graphicx} %use graph format
\usepackage{epstopdf}
\usepackage{mathrsfs}
\usepackage{makecell}
\usepackage[bookmarks=true,colorlinks,linkcolor=blue,anchorcolor=blue,citecolor=blue,unicode]{hyperref}
\usepackage{bookmark}

\usepackage{dcolumn} % 添加到导言区
\usepackage{todonotes}
\usepackage{ragged2e}
\usepackage[font=small]{caption}
\hypersetup{CJKbookmarks=true}%
\begin{document}
% \begin{CJK*}{GBK}{Song} % Use default fonts from CJK (see below)

\begin{frontmatter}
%\newpage
% \title{Connectedness resistance of the international food supply network to disturbances and shocks}
\title{Structural robustness of the international food supply network under external shocks and its determinants}

\author[SB,RCE]{Han-Yu Zhu}
\author[ECNU]{Yin-Ting Zhang}
\author[SB,RCE]{Wen-Jie Xie\corref{cor1}}
\ead{wjxie@ecust.edu.cn} 
\author[SB,RCE,Math]{Wei-Xing Zhou\corref{cor1}}
\ead{wxzhou@ecust.edu.cn} 
\cortext[cor1]{Corresponding authors.}% Address: 130 Meilong Road, P.O. Box 114, School of Business, East China University of Science and Technology, Shanghai 200237, China, Phone: +86-15216879656.}
 %to: 130 Meilong Road, P.O. Box 114, School of Business, East China University of Science and Technology, Shanghai 200237, China.}
\address[SB]{School of Business, East China University of Science and Technology, Shanghai 200237, China}
\address[RCE]{Research Center for Econophysics, East China University of Science and Technology, Shanghai 200237, China}
\address[ECNU]{School of Economics and Management, East China Normal University, Shanghai 200062, China}
\address[Math]{School of Mathematics, East China University of Science and Technology, Shanghai 200237, China}

\begin{abstract}

The stability of the global food supply network is critical for ensuring food security. This study constructs an aggregated international food supply network based on the trade data of four staple crops and evaluates its structural robustness through network integrity under accumulating external shocks. Network integrity is typically quantified in network science by the relative size of the largest connected component, and we propose a new robustness metric that incorporates both the broadness $p$ and severity $q$ of external shocks. Our findings reveal that the robustness of the network has gradually increased over the past decades, punctuated by temporary declines that can be explained by major historical events. While the aggregated network remains robust under moderate disruptions, extreme shocks targeting key suppliers such as the United States and India can trigger systemic collapse. When the shock broadness $p$ is less than about 0.3 and the shock severity $q$ is close to 1, the structural robustness curves $S(p,q)$ decrease linearly with respect to the shock broadness $p$, suggesting that the most critical economies have relatively even influence on network integrity. Comparing the robustness curves of the four individual staple foods, we find that the soybean supply network is the least robust. Furthermore, regression and machine learning analyses show that increaseing food (particularly rice and soybean) production enhances network robustness, while rising food prices significantly weaken it. 

% These results underscore the importance of international cooperation, supply chain diversification, risk management strategies, and optimizing production, storage policies, and price stability strategies in safeguarding global food security. Future research could incorporate production and stock data into network construction, refine shock simulation models, and explore the long-term adaptive capacity of the food supply network.

\end{abstract}


\end{frontmatter}

%\tableofcontents

\section{Introduction}

Global food security is broadly defined as ensuring access to sufficient, safe, and nutritious food for all people at all times \citep{Simon-1996-FoodPolicy}. However, global food security faces a multitude of challenges and uncertainties. Recent data from {\textit{The State of Food Security and Nutrition in the World 2024}} highlight the alarming trends of rising hunger, food insecurity, and malnutrition across many nations. A growing global population further exacerbates pressures on food systems, with food demand projected to rise significantly in the coming decades \citep{Godfray-Beddington-Crute-Haddad-Lawrence-Muir-Pretty-Robinson-Thomas-Toulmin-2010-Science}. These challenges are compounded by drivers such as climate change, conflict, and pandemics, which severely undermine food security worldwide. Climate change, for instance, brings extreme weather events, posing one of the greatest threats to agricultural production. In substantial regions of the global breadbaskets, climate variability explains more than 60\% of the observed yield variability \citep{Ray-Gerber-MacDonald-West-2015-NatCommun}. Conflicts, such as the Russia-Ukraine crisis, further destabilize food supply chains, intensifying food crises \citep{BenHassen-ElBilali-2022-Foods}. Meanwhile, the COVID-19 pandemic disrupted food systems globally, affecting supply, access, utilization, and stability, thereby amplifying food insecurity \citep{Laborde-Martin-Swinnen-Vos-2020-Science}. While population growth and climate change have direct impacts on food production, global events like conflicts and epidemics mainly affect food security by disrupting food supply chains \citep{Zhou-Lu-Xu-Yan-Khu-Yang-Zhao-2023-ResourConservRecycl}.

The concept of food security emphasizes not only current access to food but also the vulnerability of food systems to future disruptions \citep{Barrett-2010-Science}. Food supply within economies typically depends on domestic production, reserves, and trade. Trade plays a critical role in bridging gaps between food consumption and supply, thereby alleviating shortages in food-deficit countries \citep{D'Odorico-Carr-Laio-Ridolfi-Vandoni-2014-EarthFuture}. According to the Food and Agriculture Organization (FAO) of the United Nations, global food trade has more than doubled in the past three decades, and food trade links between countries are becoming increasingly close. The food security of all countries and regions is linked to international food trade to varying degrees, and some importing countries are increasingly relying on foreign agricultural resources to ensure their food security \citep{Fader-Gerten-Krause-Lucht-Cramer-2013-EnvironResLett}. Countries worldwide have formed a complex food supply network. While these networks enhance dietary diversity and access to food in many economies, they also introduce vulnerabilities. External shocks to food trade can disrupt these networks, destabilize global markets, and deepen food insecurity in vulnerable regions \citep{Anderson-Nelgen-2012-OxfRevEconPolicy}.

This study analyzes the structural robustness of the international food supply network (iFSN) to advance its resilience assessment. Using complex network analysis, we construct a directed weighted iFSN comprising four staple crops and apply network centrality metrics to quantify the structural importance of economies. Targeted shock simulations are conducted to evaluate cascading impacts on food supply and security, thereby identifying critical economies and systemic factors that determine network robustness. By comparing the effects of different types of disruptions, this research seeks to highlight critical economies and network vulnerabilities that shape food system robustness. The findings provide actionable insights for policymakers, enabling them to design strategies and policies to bolster global food security in the era of growing compounding uncertainties.

The rest of this paper is organized as follows. Section~\ref{S2:LitRev} provides the literature review. Section~\ref{S3:Methodology} describes the data and methods. Section~\ref{S4:EmpAnal1} analyzes the shock suimulation results. Section~\ref{S5:EmpAnal2} discusses the determinants that affect network robustness. Section~\ref{S6:Conclude} concludes.

\section{Literature review}
\label{S2:LitRev}

\subsection{Food supply network}

As the significance of food security continues to grow, the food supply network has garnered increasing attention from scholars. From an international perspective, food supply is primarily reflected through food trade. Research on international food supply and trade typically revolves around two main directions: the evolution of food trade networks and the impact of trade policies, geopolitical events, and other factors on these networks.

In terms of the evolution of food supply trade networks, scholars often use various network metrics to analyze their characteristics over time. \citet{Wang-Dai-2021-Foods} constructed a network by aggregating the trade volumes of wheat, rice, and maize, and analyzed its density, average clustering coefficient, and average path length. Their findings revealed that the food trade network is becoming progressively more complex, efficient, and tightly interconnected, with global food trade growing increasingly stable. Similarly, \citet{Zhang-Zhou-2022-FrontPhysics} constructed networks for international crop trade based on the trade values of maize, rice, soybeans, and wheat, noting that the scale of these networks has expanded and is becoming more homogeneous.
Beyond the growing interconnectedness of food supply networks, researchers have also identified distinct agglomeration properties. \citet{Duan-Nie-Wang-Yan-Xiong-2022-Sustainability}, for example, constructed a grain trade network using the trade volume of all cereals and observed a significant core-periphery structure. Likewise, \citet{Li-Xiao-Wu-Li-2024-Foods} studied the rice trade network and found that the imbalance in power distribution within the network has become markedly more pronounced. Further, \citet{Torreggiani-Mangioni-Puma-Fagiolo-2018-EnvironResLett} focused on detecting the community structure in wheat, maize, and rice trade networks, revealing an increasing tendency for these networks to form well-defined clusters of trading countries.

Regarding factors influencing food supply networks, research has predominantly explored the effects of trade policies, geopolitics, and major global events. On trade policy, \citet{Alhussam-Ren-Yao-AbuRisha-2023-Agriculture-Basel} found that food trade relations strengthened following the Belt and Road Initiative (BRI), highlighting the mutually beneficial effects of these trade agreements. Similarly, \citet{Chen-Zhang-2022-Foods} examined the grain trade among BRI countries, concluding that the initiative has significantly enhanced the grain trade network. However, they also pointed out that this network is dominated by a few large countries, which undermines its resilience to external shocks.
Geopolitical factors have also had a profound impact. \citet{Zhang-Li-Zhou-2024-Foods} studied the trade network structures of wheat, maize, and rice and found that the Russia-Ukraine conflict significantly disrupted the structural stability of the international crop trade network, particularly harming wheat trade. \citet{Laber-Klimek-Bruckner-Yang-Thurner-2023-NatFood} investigated multi-layer networks and concluded that the conflict’s comprehensive impact on Ukrainian production triggered grain shortages across Europe, West Asia, and North Africa, while dietary oils also experienced substantial losses.
Regarding global events with far-reaching impacts, \citet{Zhang-Yang-Feng-Xiao-Lang-Du-Liu-2021-Foods} examined the grain trade network during the COVID-19 pandemic and found that the global external food supply faced numerous threats, including trade disruptions, price crises, and payment difficulties. They observed a 65\% increase in the risk of the Global External Grain Supply Index (RECSI) during this period.

These studies highlight the dual nature of food supply networks: while increasing interconnectedness aids in securing food for countries, it also raises the risk of vulnerability to external shocks. Furthermore, most past research has focused on specific food types or simply aggregated trade volumes, often overlooking the differences in the caloric weight and value of various crops. To address this gap, the present study proposes the construction of an aggregate network based on calorie-weighted trade data.

\subsection{Network structure stability and influencing factors}

The notion of structural stability, first introduced by \cite{Rutledge-Basore-Mulholland-1976-JTheorBiol}, encapsulates the capacity of a complex network to withstand external disturbances and shocks without undergoing significant changes. Initially formulated to describe the stability of ecosystems, this concept has since evolved alongside the emergence of complex network theory. Its applications now span diverse domains, including social, transportation, and economic networks, with stability being analyzed across multiple dimensions.  
Building on this foundation, \cite{Albert-Jeong-Barabasi-2000-Nature} distinguished between robustness and vulnerability. 
In parallel, \cite{Haimes-2009-RiskAnal} refined and systematized the concept of resilience, defining it as the ability of a system to absorb disturbances and subsequently return to its original state. Together, these frameworks provide a nuanced understanding of stability and adaptability in complex networks, fostering interdisciplinary insights.

Efforts to measure the structural stability of networks have often relied on key metrics of connectivity. Early studies employed indicators such as average path length and network diameter to assess whether nodes in a network remain interconnected under varying conditions \citep{Pu-Yang-Pei-Tao-Lan-2013-PhysicaA, Newman-2003-SIAMRev}. 
For robustness, the focus has been on evaluating a network's ability to retain connectivity and function under simulated shocks. For instance, through systematic simulations of random and targeted attacks on wheat trade networks, \cite{Ma-Li-Li-2024-IEEETransNetwSciEng} quantitatively assessed the critical node failure thresholds triggering network collapse.
Research on network vulnerability aims to pinpoint the most fragile components susceptible to disruption. \cite{Gephart-Rovenskaya-Dieckmann-Pace-Brannstrom-2016-EnvironResLett} introduced a forward shock propagation model to quantify the redistribution of trade flows under various shock scenarios, applying it to the seafood trade network to highlight vulnerable segments. Through disruption simulations of European wheat and Asian rice production networks, \cite{Puma-Bose-Chon-Cook-2015-EnvironResLett} and \cite{Yu-Ma-Wang-2024-IntJProdEcon} systematically quantified the vulnerability of critical agricultural and semiconductor supply chains.
In the domain of resilience, In the domain of resilience research, scholars have developed diverse methodological frameworks to assess supply network resilience \citep{Poo-Wang-Yang-2024-TranspResPtA-PolicyPract,Li-Zobel-2020-IntJProdEcon}. \cite{Liang-Yu-Kharrazi-Fath-Feng-Daigger-Chen-Ma-Zhu-Mi-Yang-2020-NatFood} revealed a concerning decline in the resilience of China’s phosphorus cycle, indicating heightened vulnerability in the nation’s food system to external shocks. 
This study primarily assesses the stability of the food supply network by simulating the impact of disruptions, thereby using robustness as the key metric.

Numerous scholars have explored the factors that shape the stability of various supply chains and supply networks \citep{Brusset-Teller-2017-IntJProdEcon, Huang-Wang-Lee-Yeung-2023-IntJProdEcon, Turkes-Stancioiu-Marinescu-2024-JInnovKnowl}. In the context of food supply networks, these factors can be broadly categorized into internal and external influences.
Internal factors pertain to the structural properties of the network itself, including the degree distribution of nodes, the importance of individual nodes, and the network's efficiency and redundancy. These features play a critical role in determining the network's resilience to disruptions \citep{Li-Wang-Kharrazi-Fath-Liu-Liu-Xiao-Lai-2024-FoodSecur}.  
External factors encompass influences such as climate, transportation infrastructure, and geopolitical conditions. For instance, \cite{Tirado-Cohen-Aberman-Meerman-Thompson-2010-FoodResInt} highlighted the significant impact of climate change on food security. Similarly, \cite{Nagurney-Hassani-Nivievskyi-Martyshev-2024-EurJOperRes} underscored the crucial roles of production capacity and transportation efficiency, while \cite{Wang-Ma-Yan-Chen-Growe-2023-Foods} examined food trade networks in the context of escalating trade frictions and frequent export bans, emphasizing the importance of a stable geopolitical environment and enduring trade relationships.  

To evaluate the stability of the food supply network, this study employs a robustness index as the primary metric. The methodology involves assessing the importance of individual nodes and conducting targeted shock simulations at varying severity levels to examine how the network responds to disruptions. Additionally, quantitative models are utilized to explore the influence of various factors on network stability.

This research makes several contributions. First, it introduces an approach to constructing an aggregated food supply network by combining data on four staple foods, weighted by caloric content. This method offers a more rational basis for network construction compared to conventional approaches that simply sum trade volumes or values. Second, in measuring network robustness, this study incorporates strategic shocks specifically targeting critical nodes with varying severity levels and introduces an innovative three-dimensional framework for quantification, offering a more comprehensive evaluation of structural stability. Furthermore, this study investigates the determinants of network structural robustness, with a primary focus on external influences, including climatic conditions, production dynamics, transportation systems, geopolitical risks, and price fluctuations.  Finally, the findings are leveraged to propose policy recommendations aimed at enhancing the 
robustness of the food supply network and safeguarding global food security.

\section{Data and methodology}
\label{S3:Methodology}

\subsection{Data and iFSN construction}

Wheat, rice, maize, and soybeans are among the most crucial crops supplying calories to the global population, forming the backbone of human diets worldwide \citep{D'Odorico-Carr-Laio-Ridolfi-Vandoni-2014-EarthFuture}. To investigate the dynamics of global food supply, we use bilateral food trade data from 1986 to 2022 provided by the Food and Agriculture Organization of the United Nations (FAO, http://www.fao.org).

\begin{figure}[h!]
    \centering
    \includegraphics[width=0.8\linewidth]{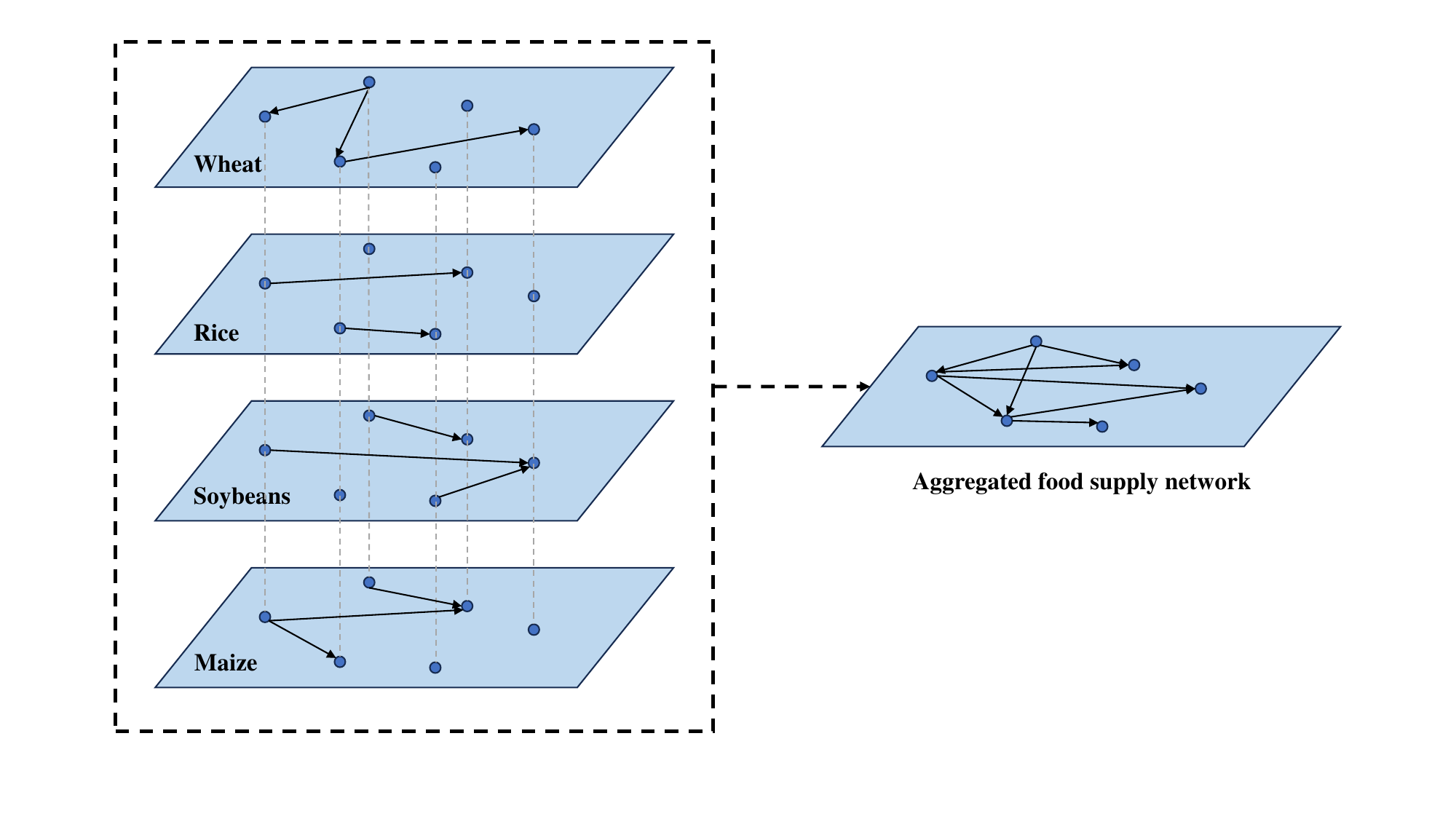}
        \vspace{-4mm}
    \caption{The construction of an aggregated food supply network. }
    \label{Fig:Aggregated_Food_supply_Network}
\end{figure}

We constructed an international food supply network (iFSN) where economies are represented as nodes. Establishing edges in such a network demands careful consideration, as directly aggregating trade volumes or monetary values is insufficient due to variations in weight and value across different crops. To address this, we calculated calorie supply flows by multiplying trade volumes by the calorie content per 1,000 tons for each crop. These flows are then aggregated across the four staples using the methodology illustrated in Fig.~\ref{Fig:Aggregated_Food_supply_Network} to form the composite iFSN.

\begin{figure}[h!]
    \centering
    \includegraphics[width=0.45\linewidth]{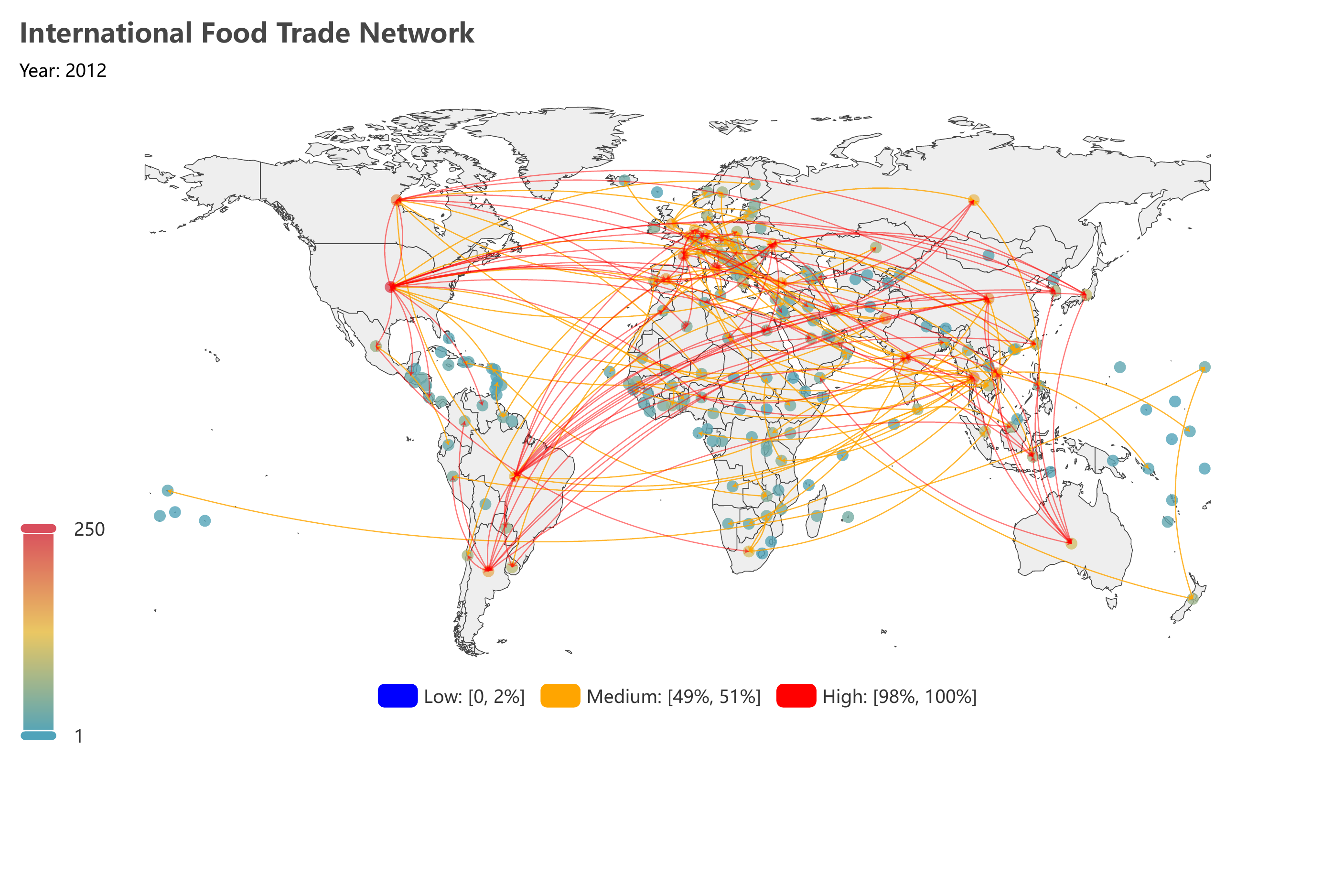}
    \includegraphics[width=0.45\linewidth]{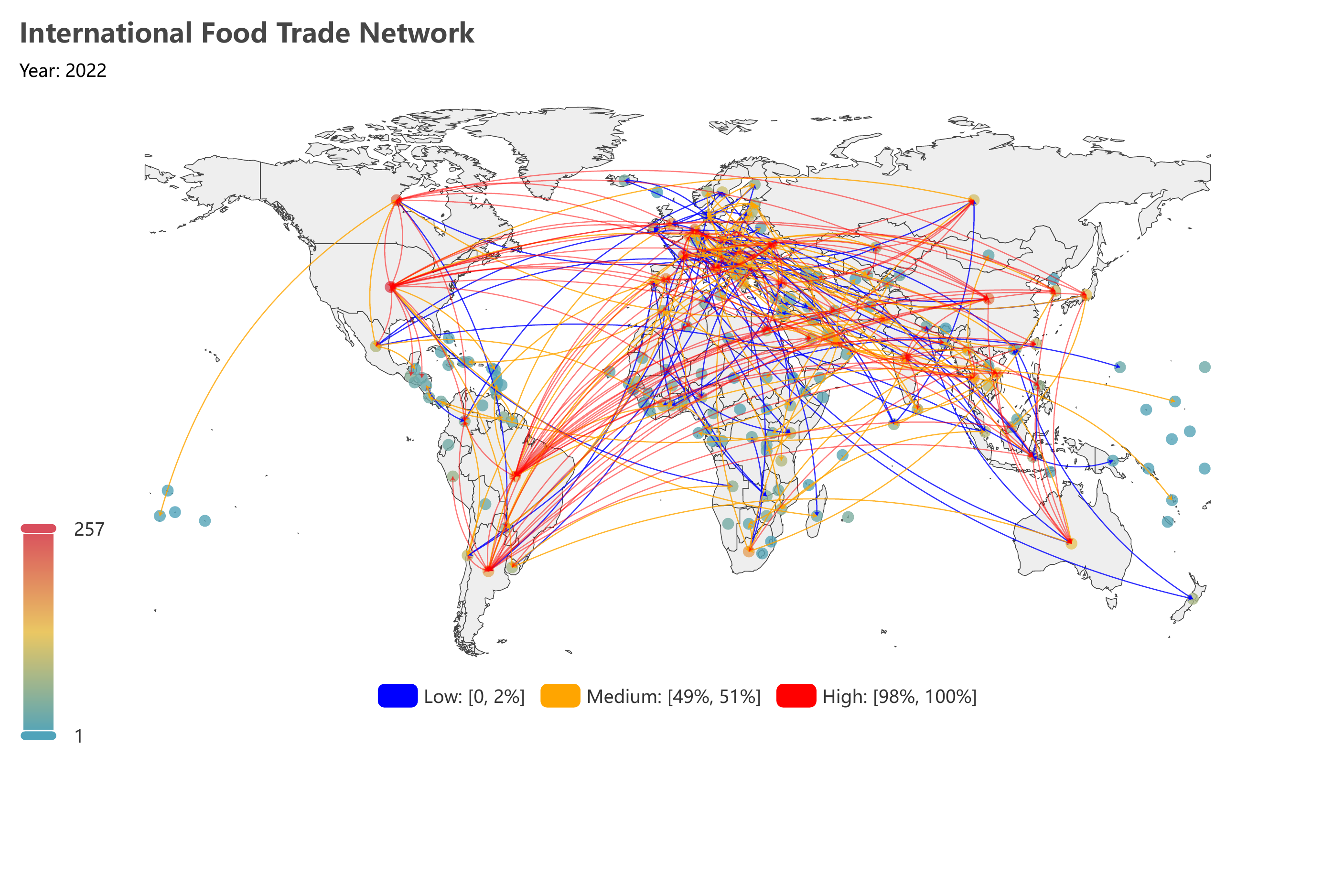}
        \vspace{-4mm}
    \caption{International food supply network in 2012 and 2022.}
    \label{Fig:Food_Supply_Network}
\end{figure}

Fig.~\ref{Fig:Food_Supply_Network} presents the global food supply networks for 2012 and 2022. Nodes represent economies, while directed links indicate food supply relationships. Node colors correspond to the number of trading partners each economy has. To enhance clarity amidst the network’s complexity, the visualization highlights edges within the upper 2\%, 49–51\%, and lower 2\% quantiles of total network weight. Notably, in 2022, the lower 2\% quantile of the food supply network contains more edges than in 2012, reflecting the growing emphasis on food security and the diversification of supply chains across economies.

\begin{figure}[h!]
    \centering
    \includegraphics[width=0.6\linewidth]{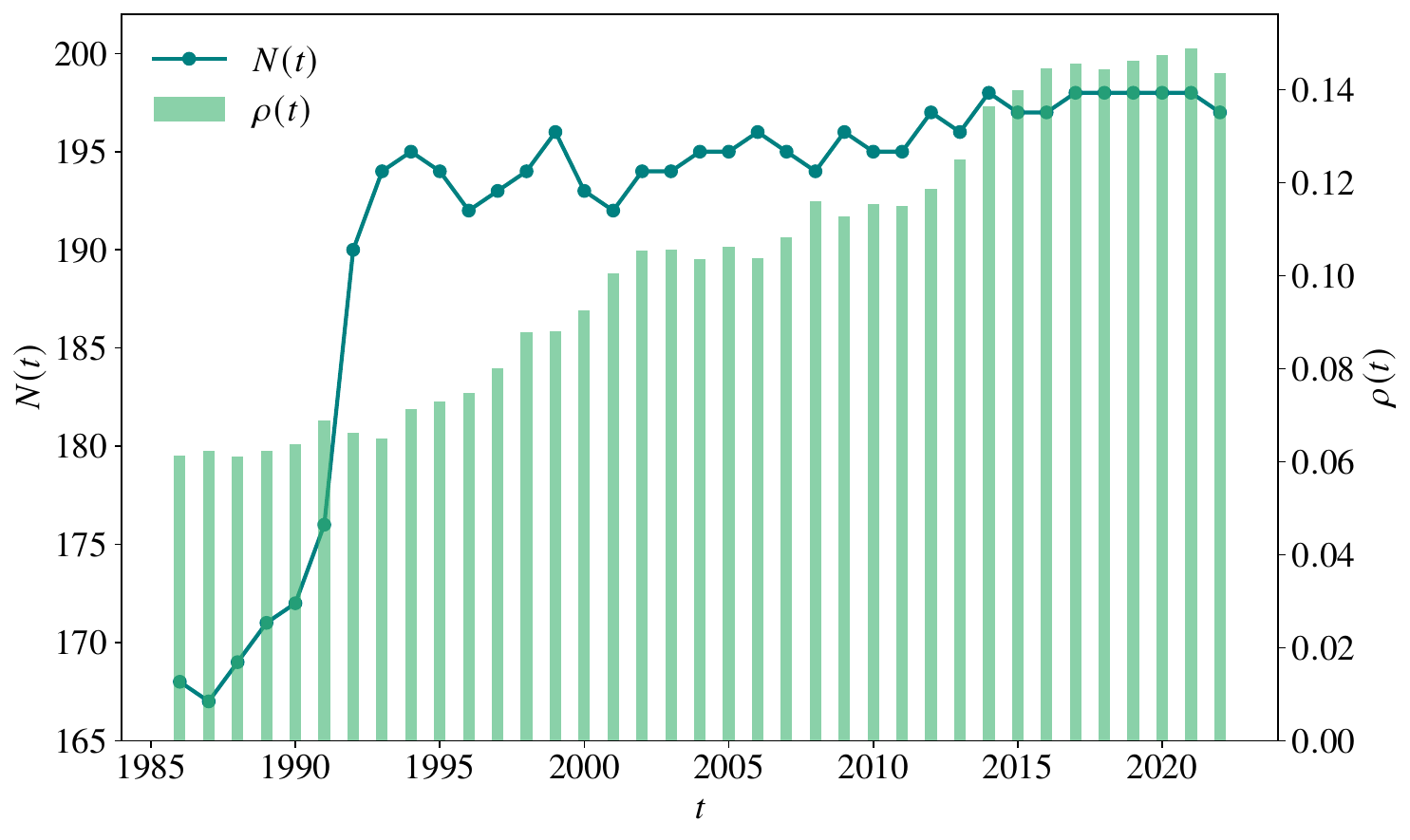}
    \caption{Evolution of the number of economies $N$ and network density $\rho$ from 1986 to 2022.}
    \label{Fig:Network_Evolution}
\end{figure}

To examine the temporal evolution of the food supply network, we calculated the number of nodes, 
$N$, and the network density, $\rho = \frac{N_E}{N(N-1)}$, as shown in Fig.~\ref{Fig:Network_Evolution}, where $N_E$ is the number of edges and $N(N-1)$ is the maximum possible number of edges in a directed network. The results reveal a steady expansion in the number of economies participating in the network, with a sharp increase in 1991, likely linked to the disintegration of the Soviet Union. Following this period of rapid growth, the number of economies stabilized in 2016 after several fluctuations. Network density, a measure of interconnectedness, exhibited continuous growth from 1986 to 2022, underscoring the increasing integration of global food systems. However, a notable decline in density occurred in 2022, which may be attributable to disruptions caused by the Russia-Ukraine conflict.

\subsection{Economic influence measurement in food supply networks}

In food supply networks, disruptions to economies of varying importance affect the network differently. Various approaches exist for quantifying the significance of individual economies within these networks \citep{Hao-Manuel-Matus-Zhang-Zhou-2009-PhysRep}. In this study, we employ four distinct types of indicators \citep{Zhang-Zhou-2023-ChaosSolitonsFractals}: those based on local network structure, global network structure, network modularity, and information theory.

\subsubsection{Measuring economic influence based on the local structure of food supply network}

Measures based on local network structure typically assess a node’s influence by considering only its immediate neighbors. The most widely used centrality measures in this category are degree and local clustering coefficient. There are also other measures of economic trade influence based on the local structure of the network in directed networks, such as semi-local centrality \citep{Chen-Lu-Shang-Zhang-Zhou-2012-PhysicaA}. Here, we use degree and local clustering coefficient to evaluate the influence of economies within the network.

For directed networks, we define in-degree centrality (ID) and out-degree centrality (OD):
\begin{equation}
    ID_{i} = \frac{k_{i}^{\text{in}}}{N - 1}
\label{Eq:ID}
\end{equation}
\begin{equation}
    OD_{i} = \frac{k_{i}^{\text{out}}}{N - 1}
\label{Eq:OD}
\end{equation}
where $k_{i}^{\text{in}}$ is the in-degree of node $i$ and $N$ is the total number of nodes in the network.

The local clustering coefficient (CC) of economy $i$ is the ratio of the number of actual trade relations between it and its neighboring economies to all possible trade relations \citep{Watts-Strogatz-1998-Nature}. It is defined as follows:
\begin{equation}
CC_{i} = \frac{\left| \left\{ a_{jk} : j, k \in \mathcal{N}_i, \, a_{jk} \in A \right\} \right|}{k_i (k_i - 1)},
\label{Eq:CC}
\end{equation}
where $\mathcal{N}_i$ is the set of neighboring economies of economy $i$ and $k_i$ denotes the size of the $\mathcal{N}_i$.

\subsubsection{Measuring economic influence based on the global structure of food supply network}

The influence of an economy within a network is intrinsically linked to the global structure of that network, so it is necessary to measure the influence of an economy based on the global structure of the network. This study employs several centrality measures---betweenness centrality (BC), closeness centrality (IC\&OC), PageRank (PR), as well as Authority (AU) and Hub (HU) metrics---to quantify the influence of economies in the context of the global food supply network.

The betweenness centrality of an economy is defined as the ratio of the number of paths passing through the economy to the total number of shortest paths \citep{Freeman-1977-Sociometry}. This metric quantifies the capacity of economy 
$i$ to influence food trade along the shortest routes connecting paired economies. A higher betweenness centrality indicates greater significance of the economy within the network. The calculation is as follows:
\begin{equation}
   BC_i = \sum_{\substack{j, k \in \mathcal{V}, j \neq k \neq i}} \frac{h_{jk}^i}{g_{jk}},
   \label{Eq:BC}
\end{equation}
where $h_{jk}^i$ refers to the number of shortest paths from node $j$ to node $k$ passing through $i$, and $g_{jk}$ refers to the total number of all shortest paths from $j$ to $k$.

Closeness centrality measures the proximity of a node to all other nodes in the network, providing insights into the efficiency of food supply transfers from a given economy to others. A higher closeness centrality indicates that the economy is positioned closer to the center of the network. In a directed network, we define inbound closeness centrality (IC):
\begin{equation}
IC_i = \frac{D_i^{\text{in}}}{N - 1} \frac{D_i^{\text{in}}}{\sum_{j=1}^{D_i^{\text{in}}} d_{ji}}
\label{Eq:IC}
\end{equation}
and outbound closeness centrality (OC):
\begin{equation}
   OC_i = \frac{D_i^{\text{out}}}{N - 1} \frac{D_i^{\text{out}}}{\sum_{j=1}^{D_i^{\text{out}}} d_{ij}}
   \label{Eq:OC}
\end{equation}
where $D_i^{\text{in}}$ and $D_i^{\text{out}}$ represent the number of nodes that can reach node $i$ and the number of nodes that can be reached by node $i$, respectively, and $d_{ji}$ and $d_{ij}$ represent the shortest path lengths from node $j$ to node $i$ or from node $i$ to node $j$, respectively.

PageRank (PR), initially developed to assess the importance of web pages, has since been widely adopted to identify nodes that significantly influence the structure of directed networks. The algorithm begins by assigning an initial importance score to each node, subsequently iterating to derive a final score based on the importance of its connected nodes \citep{Brin-Page-1998-ComputNetwISDNSyst}.The basic idea of the algorithm is that the importance of a node depends on the importance of the nodes connected to it, and the more important the node, the greater its contribution.

Authority (AU) and Hub (HU) are key metrics derived from the HITS algorithm, which, like PageRank, is designed for ranking web pages. In the HITS framework, the Hub value reflects the extent to which a node links to other significant nodes, while the Authority value indicates how much a node is linked to by these important nodes. Through iterative calculations, the HITS algorithm effectively identifies hub nodes and authoritative sources within the network \citep{Kleinberg-1999-JACM}. By calculating the Authority and Hub values, we can determine the importance of an economy in the food supply network.

\subsubsection{Measuring economic influence based on the modular structure of food supply network}

Complex networks often exhibit modular structures, characterized by a higher density of links among clustered nodes within modules compared to those between nodes in different modules \citep{Clauset-Newman-Moore-2004-PhysRevE}. Nodes that share similar roles typically display similar topological features \citep{Guimera-Amaral-2005-Nature}. To delineate these modules, we first apply the asynchronous label propagation algorithm (LPA) \citep{Raghavan-Albert-Kumara-2007-PhysRevE}. Subsequently, we employ 
within-module degree (IM) to quantify the degree of connection between a node and other nodes within its module \citep{Guimera-Amaral-2005-Nature}, while outside-module degree connectivity (OM) measures the degree of connection between a node and nodes in external modules \citep{Xu-Pan-Muscoloni-Xia-Cannistraci-2020-NatCommun}. The calculations for within-module and outside-module degree are as follows:
\begin{equation}
IM_i = \frac{\kappa_{i, m_i} - \overline{\kappa}_{m_i}}{\sigma_{\kappa_{m_i}}}
\label{Eq:IM}
\end{equation}
where $\kappa_{i, m_i}$ is the number of edges from node $i$ to other nodes in its module $m_i$, $\overline{\kappa}_{m_i}$ is the average modularity $\kappa_{i}$ of all economies in module $m_i$, and $\sigma_{\kappa_{m_i}}$ is the standard deviation of $\kappa_{i}$ in $m_i$, and
\begin{equation}
OM_i = \frac{\kappa'_{i, m_i} - \overline{\kappa}'_{m_i}}{\sigma_{\kappa_{m_i}}}
\label{Eq:OM}
\end{equation}
where $\kappa'_{i, m_i} $ is the number of edges from node $i$ to other economies outside its module $m_i$, and $\overline{\kappa}'_{m_i}$ is the average of $\kappa_{i}$ of all nodes outside $m_i$.

\subsubsection{Measuring economic influence based on information theory}

We employ mutual information to assess the importance of nodes within the network. Shannon defined information as the transfer of meaning or the quantification of uncertainty, suggesting that information serves to reduce uncertainty \citep{Borst-Theunissen-1999-NatNeurosci}. According to information theory, the information associated with each node reflects its significance, with the information content of a node being determined by its connections \citep{Liu-Jin-Zhang-Xu-2014-JSupercomput}.

In an undirected network, the mutual information of nodes is as follows:
\begin{equation}
I_{ij} = 
\begin{cases}
\ln k_i - \ln k_j, & j \in V_i \\
0, & \text{others}
\end{cases}
\label{Eq:MI1}
\end{equation}
The total mutual information for node $i$ can be calculated as follows:
\begin{equation}
M_i = \sum_{j=1}^N I_{ij}
\label{Eq:MI2}
\end{equation}

The international food supply network functions as a directed weighted network, necessitating consideration of both the direction and weight of edges when calculating the mutual information of a node. A node may serve as both a sender and a receiver. We define the probability of out-links and in-links of node $i$. The  probability of out-link of node $i$ is the probability of the edge weight from node $i$ to $j$:
\begin{equation}
p_{i}^{\text{out}} = \frac{w_{ij}}{s_i^{\text{out}}}
\label{Eq:pout}
\end{equation}
The calculation of the probability of the in-links for node $j$ follows a similar approach:
\begin{equation}
p_{j}^{\text{in}} = \frac{w_{ij}}{s_j^{\text{in}}}
\label{Eq:pin}
\end{equation}
Then the mutual information $I_{ij}^{w}$ from node $i$ to node $j$ is:
\begin{equation}
I_{ij}^{w} = 
\begin{cases}
\ln \frac{s_{i}^{\text{out}}}{w_{ij}} - \ln \frac{s_{j}^{\text{in}}}{w_{ij}}, & j \in V_i^{\text{out}} \\
0, & \text{others}
\end{cases}
\label{Eq:Iij}
\end{equation}

The total mutual information contained in node $i$ is calculated as the sum of the mutual information from node $i$ to all nodes it points to, minus the sum of the mutual information from all nodes that point to $i$:
\begin{equation}
MI_i = \sum_{j \in V_i^{\text{out}}} I_{ij}^{w} - \sum_{s \in V_i^{\text{in}}} I_{si}^{w}
\label{Eq:MI}
\end{equation}

\subsection{Shocks to nodes and structural robustness of iFSN}

From Fig.~\ref{Fig:Food_Supply_Network}, it is evident that the network contains several highly connected nodes that serve as hubs linking numerous distinct economies. These critical nodes play a central role in maintaining the network's connectivity and functionality. Given their importance, the failure of such nodes could significantly disrupt the overall system efficiency \citep{Crucitti-Latora-Marchiori-2004-PhysRevE}. Therefore, we focus on strategic shock models to evaluate network robustness. Based on the node importance assessment described earlier, we systematically apply these shocks to key nodes to analyze their impact on the network's stability and performance \citep{Albert-Jeong-Barabasi-2000-Nature}.

First, we discussed the simulation of node shocks. To simulate the effects of node shocks, the simplest approach is to remove all edges connected to the targeted nodes. Under strategic shocks, this corresponds to severing all connections involving the selected economies. By simulating the network under various shock scenarios, we aimed to identify vulnerabilities that may lead to network collapse, enabling targeted interventions to enhance structural stability.

To better approximate real-world food supply disruptions, we introduced two parameters, $p$ and $q$, in our simulation framework \citep{Wei-Xie-Zhou-2022-Energy}. The parameter $p$ represents the broadness of the shock, defined as the proportion of economies under shock relative to the total number of economies. For instance, $p$ indicates that the proportion of $p$ economies are targeted for shock, ranked by their importance. This parameter reflects the varying scales of real-world disruptions. For example, during the Sino-US trade friction that began in 2018, targeted trade restrictions led to localized disruptions, while the COVID-19 pandemic in 2019 caused widespread transportation constraints, resulting in global impacts.
The second parameter, $q$, quantifies the severity level of the shock, defined as the proportion of links severed for each targeted economy. In real-world scenarios, economies typically have some level of resistance to shocks, making it unrealistic to assume that all links are severed during a shock. By introducing $q$, we simulated scenarios in which a proportion $q$ of an economy's links are randomly removed, capturing varying degrees of network damage. Together, the parameters $p$ and $q$ enable us to explore the effects of shocks of differing scales and severity levels on the food supply network.

The robustness of the food supply network can be conceptualized as its ability to maintain functional connectivity when some nodes are disrupted. Within the framework of food security, this corresponds to whether other economies can effectively compensate for disruptions in food supply. We quantify network robustness by measuring connectivity, specifically the proportion of nodes in the largest connected subgraph after a given shock, denoted as $S(p)$.
To further evaluate the network's structural stability under varying shock scenarios, we define robustness $R$ as a comprehensive measure of network robustness \citep{Schneider-Moreira-Andrade-Havlin-Herrmann-2011-ProcNatlAcadSciUSA}:
\begin{equation}
R_{p} = \frac{1}{N} \sum_{n=1}^{N} S\left(\frac{n}{N}\right) = \frac{1}{N} \sum_{n=1}^{N} S(p)
\label{Eq:Robustness}
\end{equation}
where $N$ represents the total number of economies in the international food supply network and $n = Np$. The normalization factor ensures that the robustness metric is comparable across networks of different sizes.
When nodes experience shocks---whether due to climate shocks, geopolitical conflicts, or trade disputes---other nodes may be forced to abandon existing food supply relationships, either voluntarily or passively. This leads to a reduction in $S(p)$. Consequently, the stability of the network structure deteriorates.

\begin{figure}[h!]
    \centering
    \includegraphics[width=0.4\linewidth]{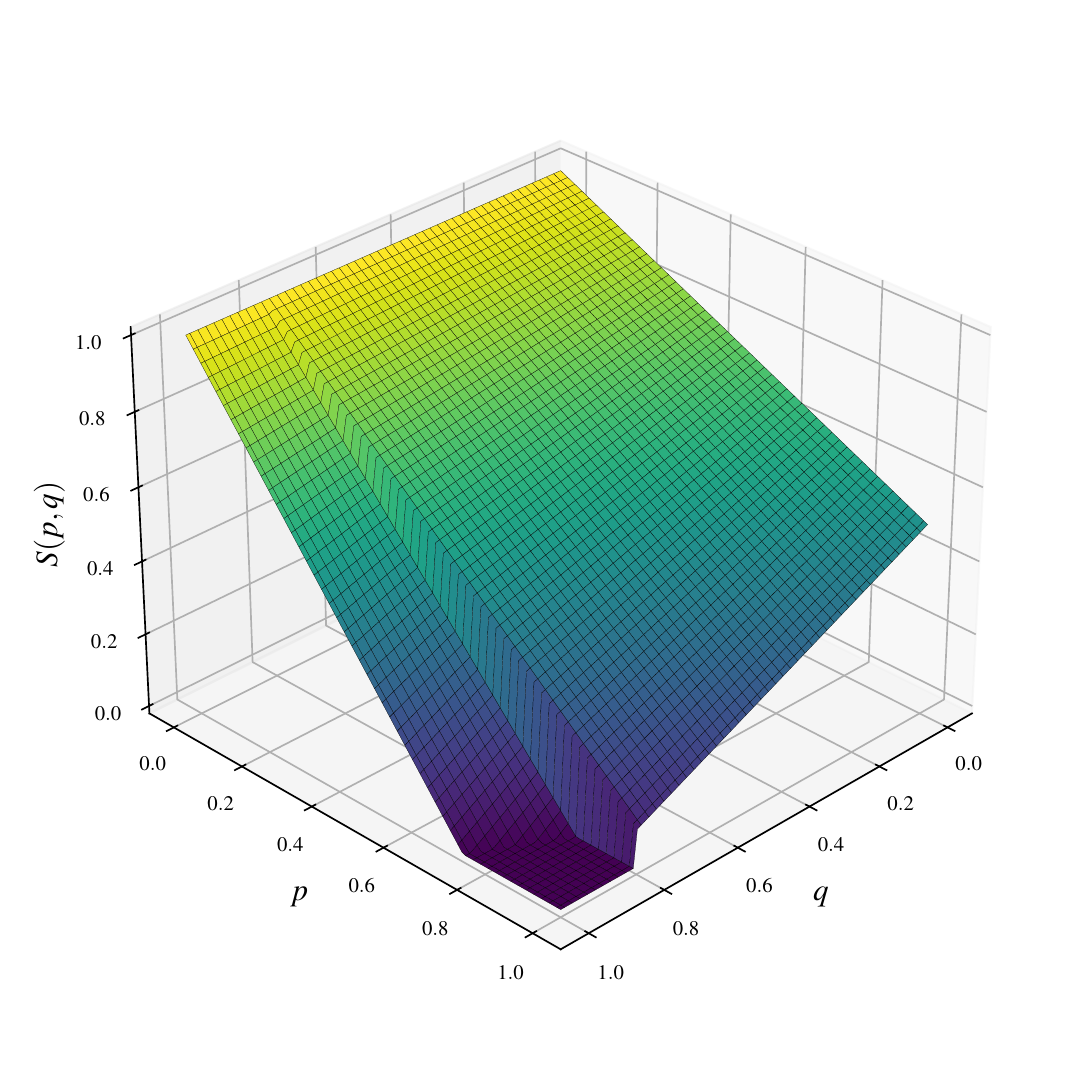}
    \caption{Structural robustness demonstration}
    \label{Fig:Demonstration}
\end{figure}

We assume that the structural robustness of a given network remains invariant, irrespective of the shock range or severity. To explore this, we define a three-dimensional coordinate system, where $p$, $q$, and the proportion of nodes in the largest connected subgraph $S(p)$ constitute the axes. This framework is illustrated in Fig.~\ref{Fig:Demonstration}. To quantify robustness, we propose evaluating the volume enclosed by the curve within this coordinate space:
% \begin{equation}
%   R_{p, q}=\int_{q=0}^{1}\int_{p=0}^{1} S(p, q) {\rm{d}}p{\rm{d}}q
% \label{Eq:Robustness_volumn}
% \end{equation}
\begin{equation}
   R_{p, q}=\frac{1}{N}\frac{1}{M} \sum_{n=1}^{N}\sum_{m=1}^{M}S\left(\frac{n}{N},\frac{m}{M}\right) =\frac{1}{N}\frac{1}{M} \sum_{n=1}^{N}\sum_{m=1}^{M}S\left(p, q\right) 
\label{Eq:Robustness_volumn}
\end{equation}
where $S\left(\frac{n}{N},\frac{m}{M}\right)$ represents the node ratio of the largest connected subgraph when the shock broadness is $p=\frac{n}{N}$ and the shock severity is $q=\frac{m}{M}$.
$M$ represents the number of divisions of the interval $[0,1]$.
 
\section{Qualifying network structural robustness}
\label{S4:EmpAnal1}

\subsection{Ranking of economic influence in the iFSN}

Node importance indicators are frequently employed to assess the role of economies in complex networks \citep{Zhang-Zhou-Long-Wang-Hong-Armaghan-2023-JKingSaudUniv-ComputInfSci, Richmond-2019-JFinanc, Wei-Xie-Zhou-2022-Energy}. The choice of indicator influences the evaluation of an economy's importance.  While many studies have utilized multiple indicators for measurement, comprehensive comparisons among these indicators remain limited, particularly in applications involving food-related complex networks. This study applies the economic influence indicators developed in Section~\ref{S3:Methodology} to evaluate the importance of economies in the food supply network across multiple years.  Taking 2022 as an example, we identify the top 10 economies for each indicator (Table~\ref{Table: Top 10}).

Each type of indicator yields unique insights into the network. Degree centrality identifies key food suppliers and demanders, with economies such as the Netherlands ranking highly in in-degree centrality due to food imports from multiple partners. 
The clustering coefficient indicator identifies economies that play a crucial role within specific regions. The top ten economies are predominantly island nations and African countries, whose food supply models are characterized by close ties to a limited number of supply partners and a strong regional dependency.
Centrality measures such as betweenness centrality underscore economies like the United States, Canada, and India as pivotal hubs in the global network. The top three economies in inbound closeness centrality are the Netherlands, the United States, and Canada, mainly because these three economies have more food supply partners. The top three  economies in outbound closeness centrality are India, the United States, and Thailand. India and Thailand mainly export rice, while the United States mainly exports maize and soybeans.
Algorithm-derived indicators elucidate influential nodes based on food supply, transit, and demand. Meanwhile, module-specific indicators capture economies influential within or across regional clusters. For instance, the United States and India exhibit high intra- and inter-module connectivity. The U.S. links North America with Europe and Asia through exports of soybeans, wheat, and maize, while India connects Southeast Asia with the Middle East and Africa, primarily through rice exports. Information-theoretic measures reveal the United States, Canada, and India as critical to balancing supply-demand dynamics in the network.

A comparative analysis of the top 10 economies across indicators reveals key patterns. First, strong alignment exists between import- and export-based rankings, reflecting their shared focus on food trade flows. Second, intra- and inter-module connectivity rankings often coincide, highlighting economies significant within regions that also serve as cross-regional hubs. Third, comprehensive indicators like betweenness centrality and mutual information demonstrate a high degree of overlap. Conversely, clustering coefficient results differ significantly from algorithm-based measures, emphasizing their regional specificity.

We counted the frequency of each country appearing in Table~\ref{Table: Top 10}, as shown in Table~\ref{Table: Top 10 freq}. We found that the United States is the most frequently ranked economy, underscoring its critical role as a soybean and maize exporter. India, China, and France follow, with India's importance stemming from its status as the largest rice exporter. France's influence is enhanced by its central position in the European food supply network and its role as a major wheat exporter. Other significant exporters, including Canada and Argentina, owe their prominence to mechanized farming and high-yield agricultural systems, primarily supplying wheat and soybeans, respectively. We use “others” to denote the remaining economies that appear only once.

Most frequently ranked economies are major exporters, highlighting the criticality of production-heavy nodes in the food supply network. Furthermore, the overlap in rankings across indicators suggests underlying correlations between measures, meriting further investigation.

\begin{table}[htbp]
\centering
\caption{Top 10 economies identified based on different influence indicators in 2022.}
\begin{tabularx}{\textwidth}{lXXXXXXXXXXXX}
\toprule
$ID$   & $OD$   & $CC$   & $BC$   & $IC$   & $OC$   & $PR$   & $HU$   & $AU$   & $IM$   & $OM$   & $MI$   \\
\midrule
NLD  & IND  & FSM  & USA  & NLD  & IND  & COD  & BRA  & CHN  & IND  & IND  & USA  \\
CAN  & USA  & SSD  & CAN  & USA  & USA  & RWA  & USA  & MEX  & ZMB  & CHN  & CAN  \\
USA  & THA  & TLS  & IND  & CAN  & THA  & IND  & UKR  & JPN  & USA  & USA  & IND  \\
FRA  & CHN  & GAB  & FRA  & FRA  & CHN  & NLD  & AUS  & IRN  & THA  & THA  & FRA  \\
GBR  & ARG  & LUX  & ZAF  & GBR  & ARG  & AGO  & ARG  & ESP  & TZA  & ITA  & BRA  \\
DEU  & PAK  & DMA  & ARE  & DEU  & ITA  & DEU  & CAN  & COL  & ARG  & FRA  & ARG  \\
ITA  & ITA  & COM  & NLD  & ARE  & PAK  & CHN  & FRA  & KOR  & CHN  & PAK  & NLD  \\
ARE  & TUR  & NRU  & CHN  & ITA  & TUR  & TUR  & IND  & EGY  & PAK  & ARG  & DEU  \\
BEL  & FRA  & TCD  & GBR  & ESP  & FRA  & ESP  & URY  & CAN  & ARE  & BRA  & CHN  \\
ESP  & BRA  & FRO  & TUR  & BEL  & BRA  & GBR  & RUS  & TWN  & TUR  & CAN  & THA  \\
\bottomrule
\end{tabularx}
\label{Table: Top 10}
\end{table}

\begin{table}[h!]
\centering
\caption{The counts of economies appearing in the top ten of the indicator ranking.}
\begin{tabular*}{\textwidth}{@{\extracolsep{\fill}}llcllc}
\toprule
ISO3 & Economy           & Count & ISO3 & Economy           & Count \\
\midrule
USA      & United States      & 9       & ITA      & Italy                  & 5 \\
IND      & India              & 8       & TUR      & Turkey                 & 5 \\
CHN      & China              & 8       & GBR      & United Kingdom         & 4 \\
FRA      & France             & 8       & ESP      & Spain                  & 4 \\
CAN      & Canada             & 7       & DEU      & Germany                & 4 \\
ARG      & Argentina          & 6       & PAK      & Pakistan               & 4 \\
NLD      & Netherlands        & 5       & ARE      & United Arab Emirates   & 4 \\
BRA      & Brazil             & 5       & BEL      & Belgium                & 2 \\
THA      & Thailand           & 5       & Others   & Various economies      & 1 \\
\bottomrule
\end{tabular*}
\label{Table: Top 10 freq}
\end{table}

\subsection{Correlation analysis between node importance rank indicators}

The node importance indicators are derived from the topological properties of the international food supply network, leading to potential correlations among them. In the preceding analysis, we observed that certain economies exhibited similar rankings across different node importance measures. To quantify these relationships, we computed the correlations between all node importance indicators, with the results summarized in Table~\ref{Table: correlation of indicators}.

Table~\ref{Table: correlation of indicators} presents the Spearman correlation coefficients between the rankings of node importance indicators. The analysis reveals that most indicators are significantly correlated. The highest correlations are observed between in-degree centrality and inbound closeness centrality, and between out-degree centrality and outbound closeness centrality, with coefficients approaching unity. Notably, cluster centrality exhibits significant negative correlations with other indicators, in contrast to the positive correlations observed elsewhere. This divergence stems from the nature of cluster centrality, which predominantly captures local cohesion. Nodes with high cluster centrality rankings tend to belong to tightly interconnected communities with limited external connections, whereas other indicators prioritize nodes with broader network connectivity, reflecting the principle that ``more connected nodes are more important.'' Additionally, the mutual information indicator shows relatively weak correlations with certain other measures.

The correlation coefficient results verify the similarities among the node importance indicators. However, each indicator is based on distinct theoretical foundations, making them mutually non-substitutable. Comparing these indicators across the network provides valuable insights for identifying nodes that play pivotal roles. Furthermore, simulating node shocks requires a multifaceted approach that incorporates various importance measures. This study seeks to identify critical nodes in the food supply network while assessing its structural stability under targeted disruptions. By considering shocks based on different importance indicators, we gain a deeper understanding of the factors influencing robustness and can devise more effective strategies to enhance the stability of the food supply network.

\begin{table}[htbp]
\centering
\caption{Correlation of the influence indicators of food supply economies in 2022.}
\setlength{\tabcolsep}{2mm}
\newcolumntype{d}[1]{D{.}{.}{#1}} % 定义小数点对齐列
\begin{tabular}{d{2.0} *{6}{d{6.6}}}
\toprule
 & \multicolumn{1}{c}{$ID$} & \multicolumn{1}{c}{$OD$} & \multicolumn{1}{c}{$CC$} & \multicolumn{1}{c}{$BC$} & \multicolumn{1}{c}{$IC$} & \multicolumn{1}{c}{$OC$} \\
\midrule
ID & 1 &  &  &  &  &  \\
OD & 0.759^{***} & 1 &  &  &  &  \\
CC & -0.549^{***} & -0.677^{***} & 1 &  &  &  \\
BC & 0.823^{***} & 0.912^{***} & -0.753^{***} & 1 &  &  \\
IC & 0.982^{***} & 0.752^{***} & -0.551^{***} & 0.816^{***} & 1 &  \\
OC & 0.739^{***} & 0.979^{***} & -0.639^{***} & 0.885^{***} & 0.737^{***} & 1 \\
PR & 0.596^{***} & 0.481^{***} & -0.418^{***} & 0.566^{***} & 0.616^{***} & 0.501^{***} \\
HU & 0.672^{***} & 0.927^{***} & -0.559^{***} & 0.811^{***} & 0.651^{***} & 0.919^{***} \\
AU & 0.532^{***} & 0.58^{***} & -0.402^{***} & 0.55^{***} & 0.557^{***} & 0.593^{***} \\
IM & 0.459^{***} & 0.64^{***} & -0.615^{***} & 0.608^{***} & 0.493^{***} & 0.604^{***} \\
OM & 0.459^{***} & 0.605^{***} & -0.666^{***} & 0.585^{***} & 0.489^{***} & 0.566^{***} \\
MI & 0.097 & 0.175^{*} & -0.35^{***} & 0.182^{*} & 0.098 & 0.189^{**} \\
\toprule
 & PR & HU & AU & IM & OM & MI \\
\midrule
PR & 1 &  &  &  &  &  \\
HU & 0.394^{***} & 1 &  &  &  &  \\
AU & 0.626^{***} & 0.484^{***} & 1 &  &  &  \\
IM & 0.279^{***} & 0.511^{***} & 0.282^{***} & 1 &  &  \\
OM & 0.238^{***} & 0.466^{***} & 0.3^{***} & 0.889^{***} & 1 &  \\
MI & 0.236^{***} & 0.259^{***} & 0.147^{*} & 0.311^{***} & 0.254^{***} & 1 \\
\bottomrule
\end{tabular}
    \begin{tablenotes}   
        \footnotesize        
        \item Note: * indicates significance at the 10\% level; ** at the 5\% level; *** at the 1\% level.
      \end{tablenotes}  
\label{Table: correlation of indicators}
\end{table}

\subsection{Analysis on the structure robustness of the iFSN under shocks}

As global food trade relationships continue to evolve, the structure of the international food supply network (iFSN) has become increasingly intricate. Understanding the stability of this network under trade shocks is critical for ensuring global food security. To investigate this, we applied the shock-to-node model described in Section~\ref{S3:Methodology}, simulating both random and targeted disruptions. 

\begin{figure}[h!]
    \centering
    \includegraphics[width=0.32\linewidth]{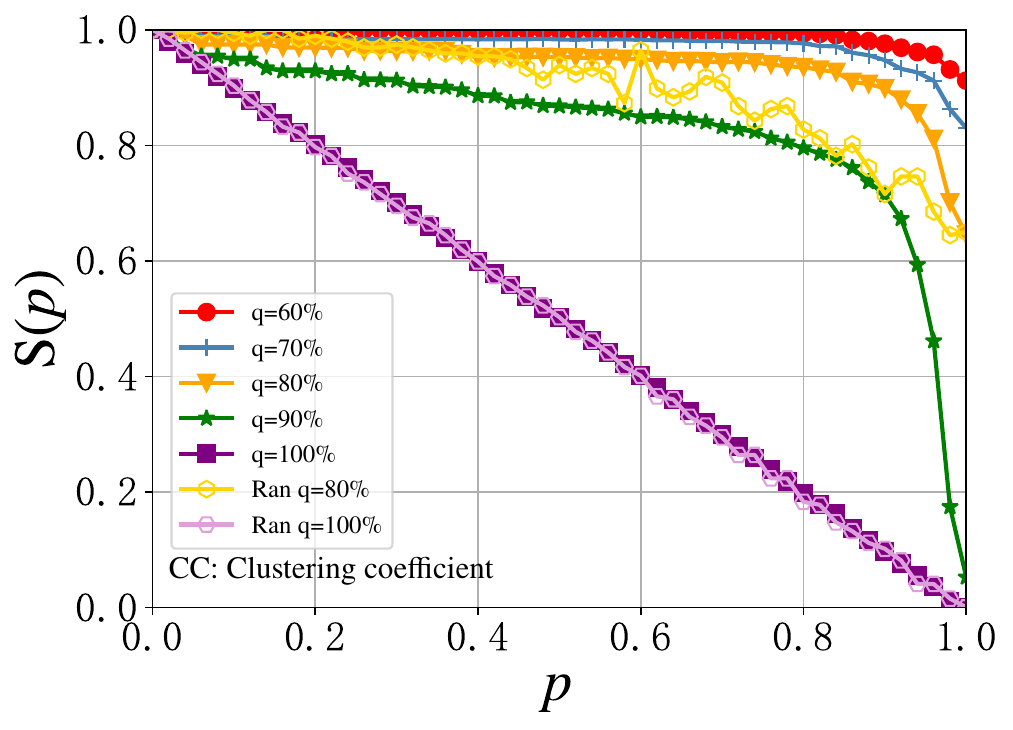}
    \includegraphics[width=0.32\linewidth]{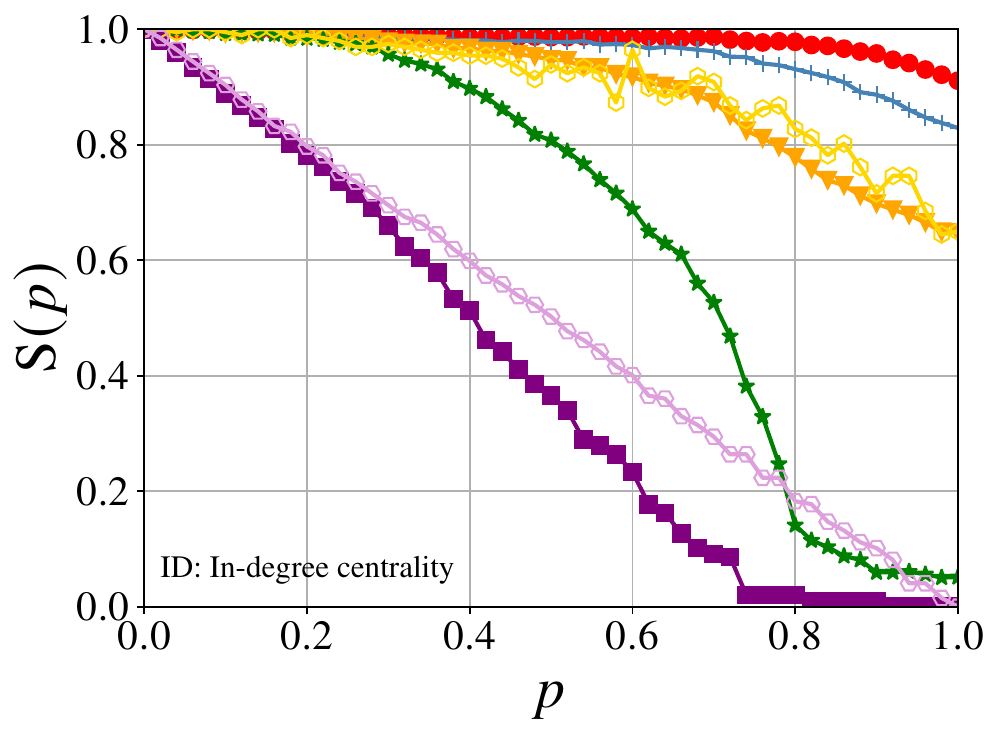}
    \includegraphics[width=0.32\linewidth]{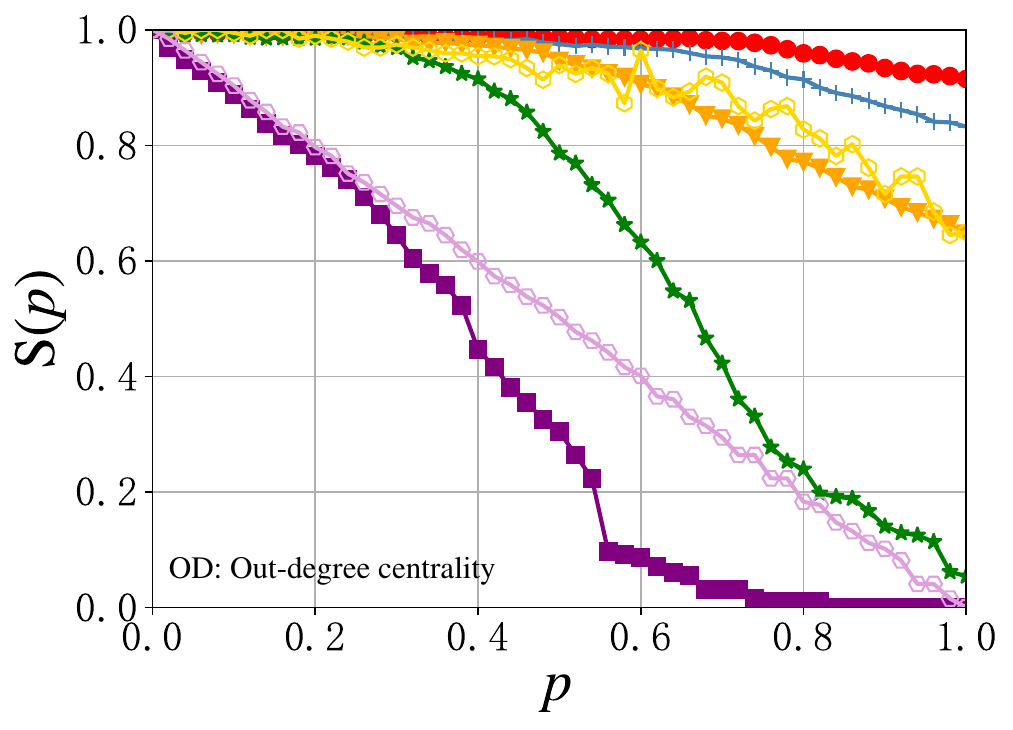}\\
    % \vspace{-3mm}
    \includegraphics[width=0.32\linewidth]{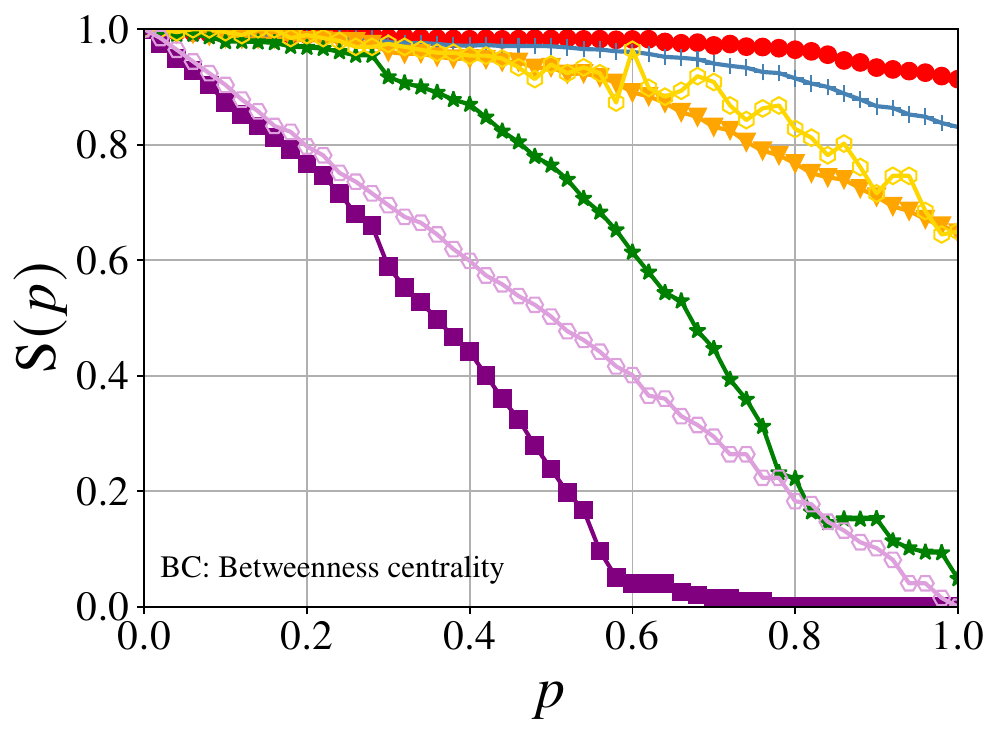}
    \includegraphics[width=0.32\linewidth]{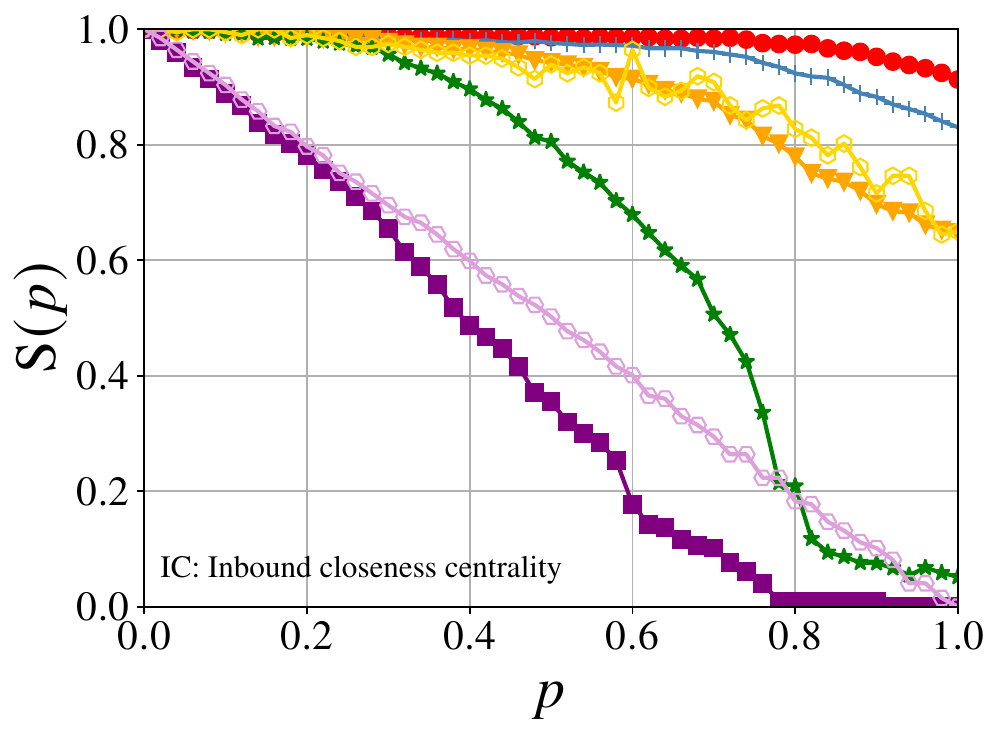}
    \includegraphics[width=0.32\linewidth]{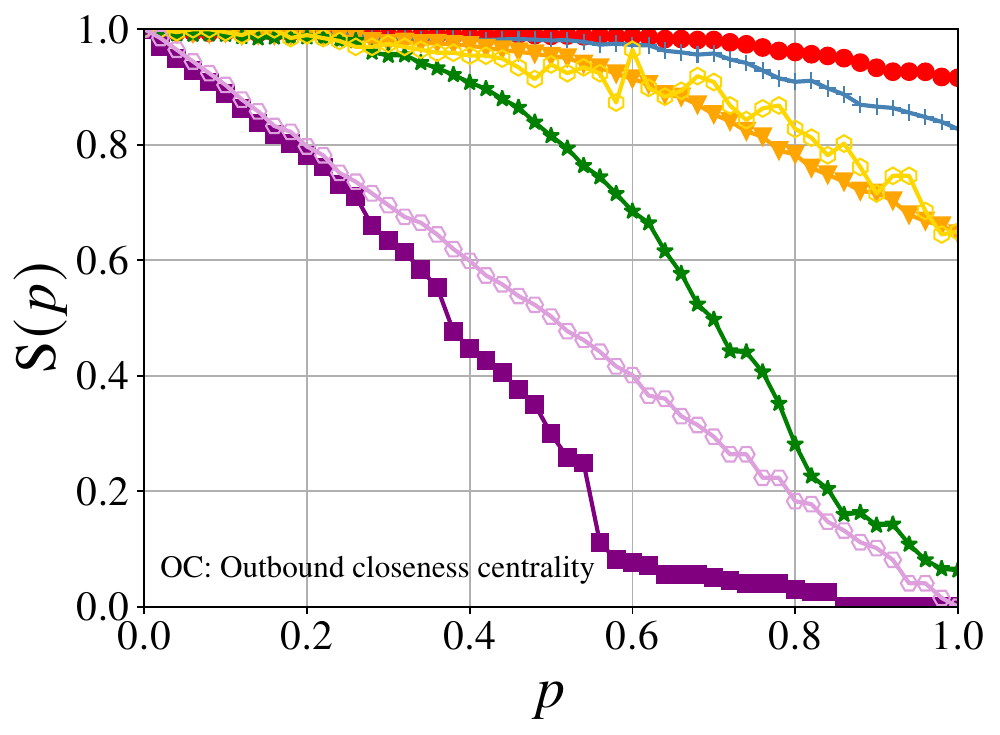}
    % \vspace{-4mm}
    \includegraphics[width=0.32\linewidth]{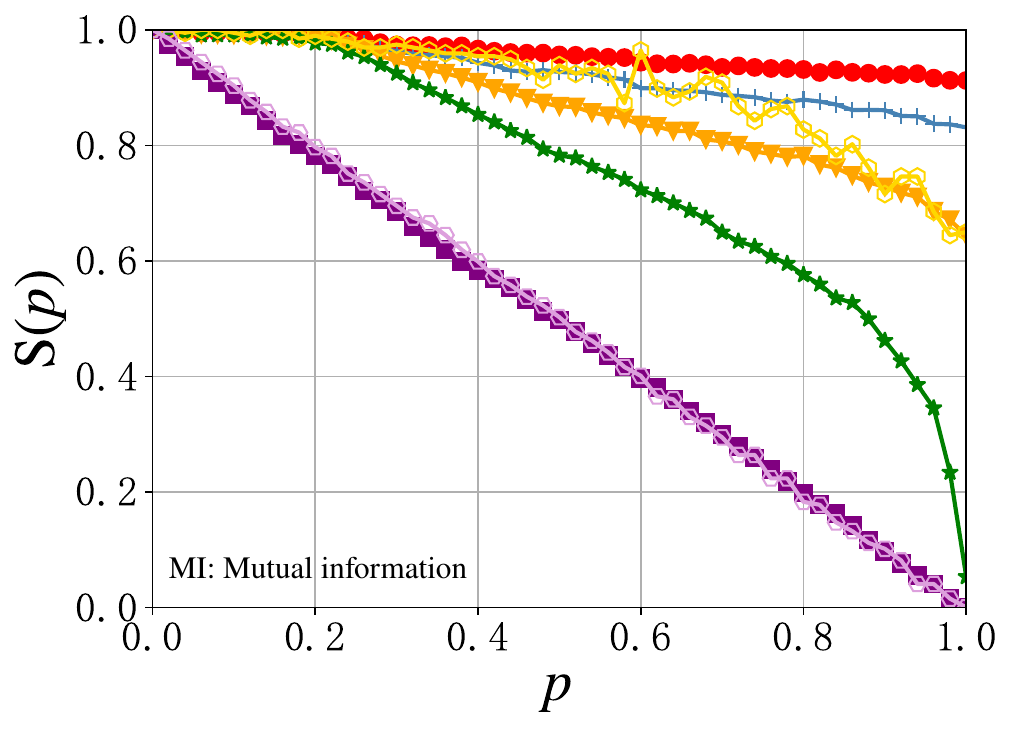}
    \includegraphics[width=0.32\linewidth]{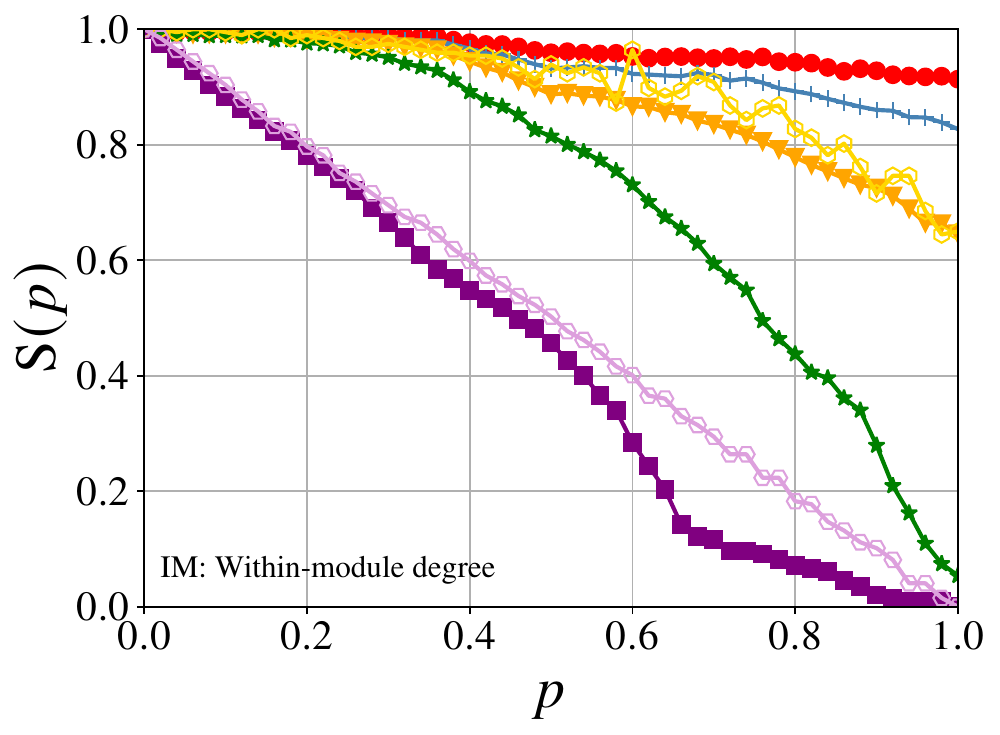}
    \includegraphics[width=0.32\linewidth]{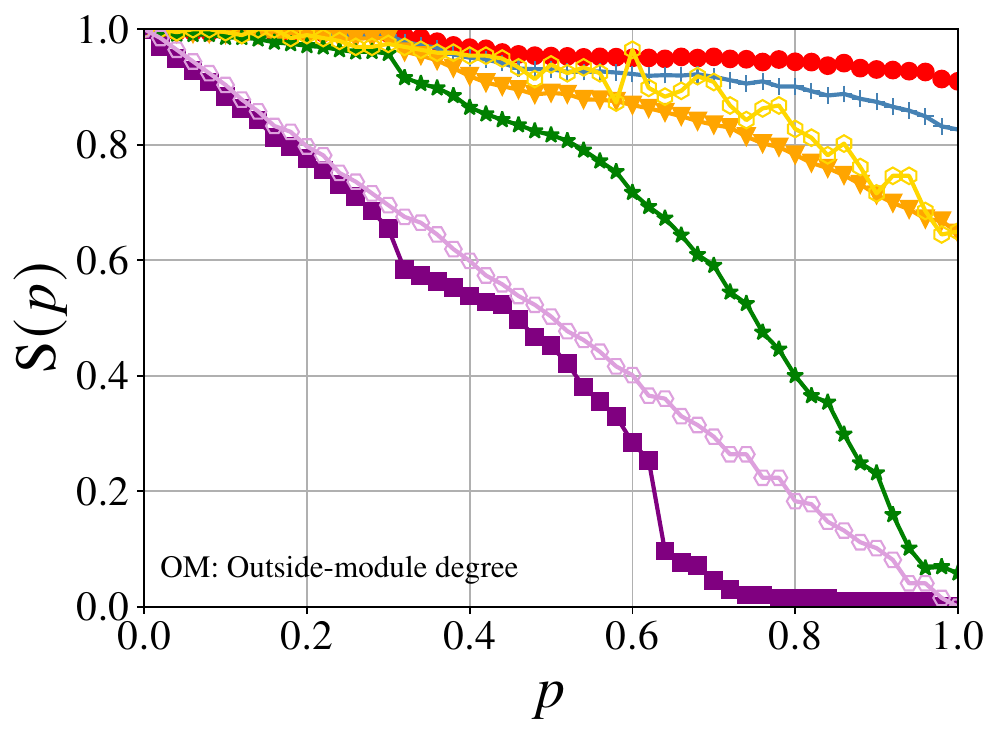}
    % \vspace{-4mm}
    \includegraphics[width=0.32\linewidth]{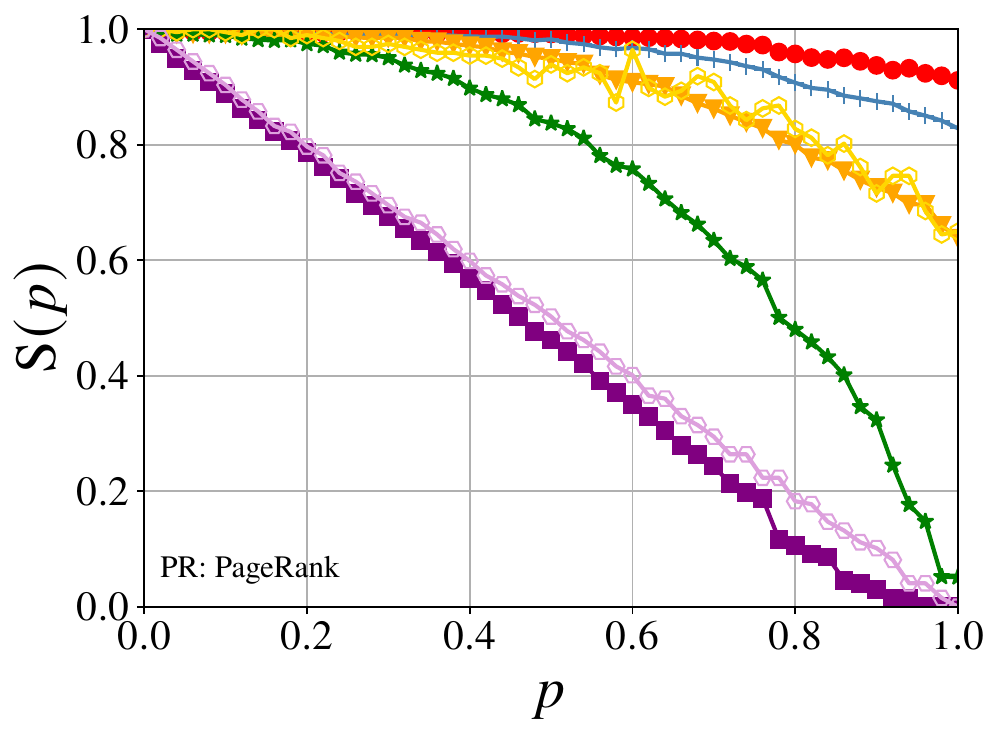}
    \includegraphics[width=0.32\linewidth]{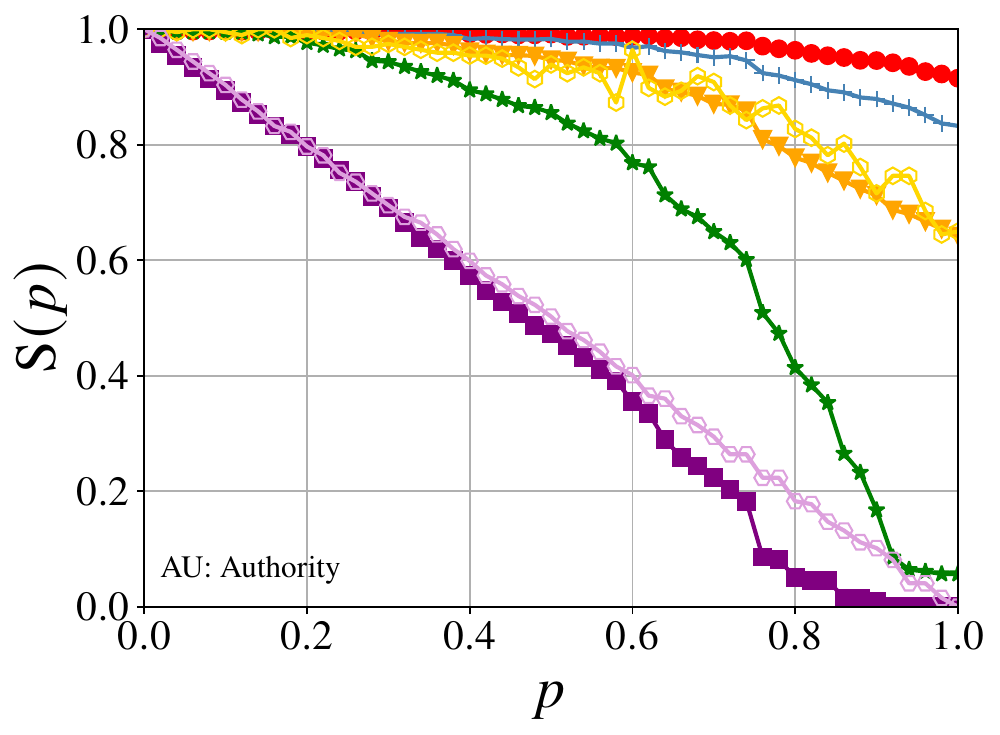}
    \includegraphics[width=0.32\linewidth]{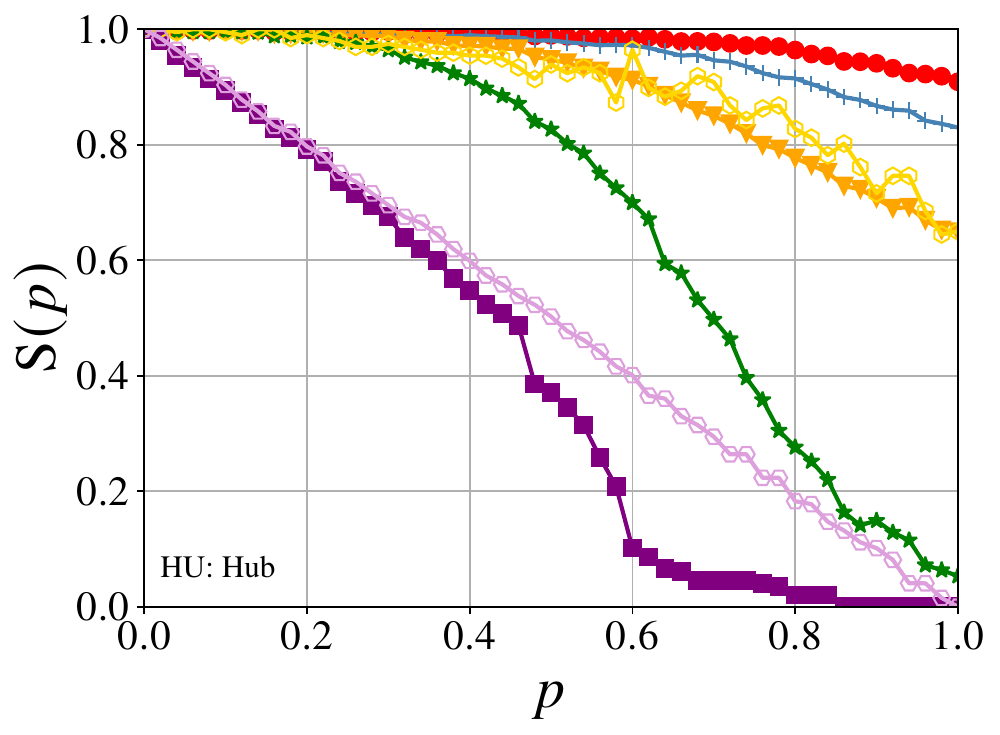}
    \caption{Structural robustness curves of the aggregated international food supply network under shocks to economies.}
    \label{Fig:NodeAttack_all}
\end{figure}

Fig.~\ref{Fig:NodeAttack_all} presents the results of focused shocks applied at severity levels $q=60\%$, $q=70\%$, $q=80\%$, $q=90\%$, and $q=100\%$, alongside random shocks at $q=80\%$ and $q=100\%$, using data from 2022. The horizontal axis represents the proportion of nodes shocked ($p$), while the vertical axis depicts network robustness, calculated via Eq.~(\ref{Eq:Robustness}). In the strategic shock simulations, connections between economies and their trading partners were removed sequentially based on economic influence rankings, affecting 2\% of economies per step.
The results indicate that the aggregated iFSN demonstrates considerable robustness under simulated disruptions. At a shock severity of 60\%, the network experiences minimal losses even when all nodes are targeted. Significant structural damage only becomes evident when the shock severity surpasses 90\%. At 90\% and 100\%, robustness declines slowly at first, followed by a sharp decrease as the proportion of shocked economies increases. This behavior underscores the high overall connectivity and redundancy of the iFSN, which is largely attributed to the diversified food supply chains of participating economies. 
Comparing targeted and random shocks, we observe that targeting influential economies accelerates network collapse, indicating that the robustness of the network is more likely to suffer greater damage. 
When examining different metrics of economic influence, we find that the damage to the network caused by economies with high clustering coefficients and mutual information under shocks is similar to that of random shocks. This suggests that these metrics are less effective for assessing the strategic importance of individual economies within the network.

Furthermore, under $q=100\%$ strategic shocks, where economies are removed based on various influence metrics, the initial decline in network robustness follows a linear trajectory with a slope corresponding to a 45-degree angle. 
This pattern indicates that influential economies are relatively evenly distributed in terms of their importance to the overall network. This phenomenon can be attributed to the aggregated nature of the network, which represents the combined trade of four staple food commodities. The diversity and complementarity across these staples contribute to the observed robustness.

\begin{figure}[h!]
    \centering
    \includegraphics[width=0.32\linewidth]{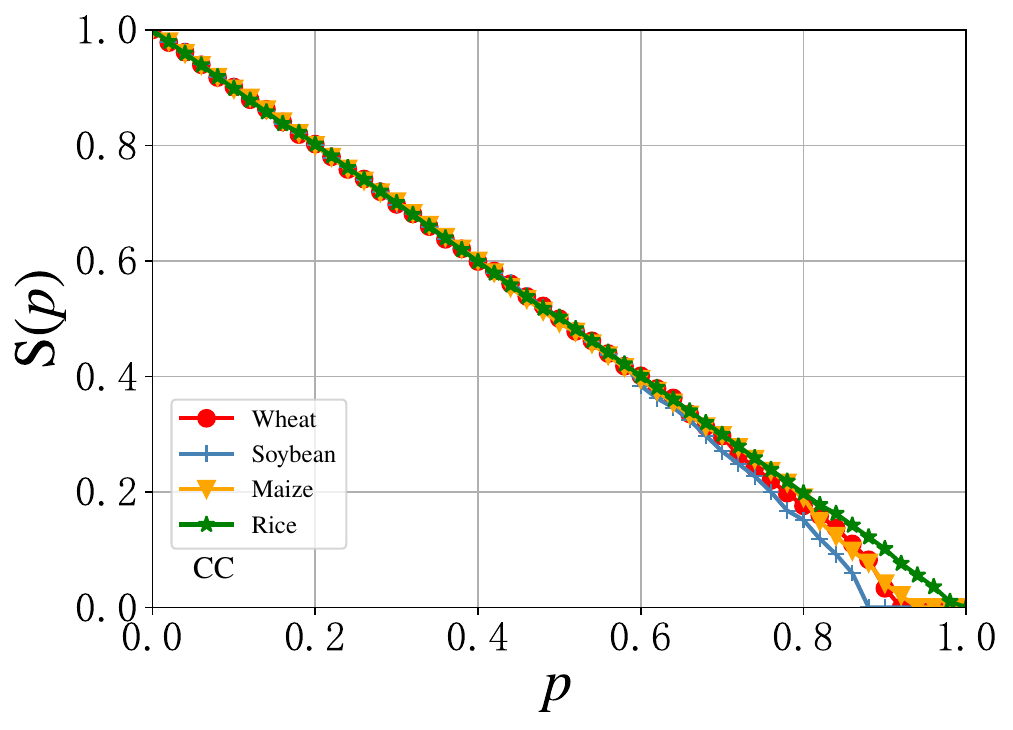}
    \includegraphics[width=0.32\linewidth]{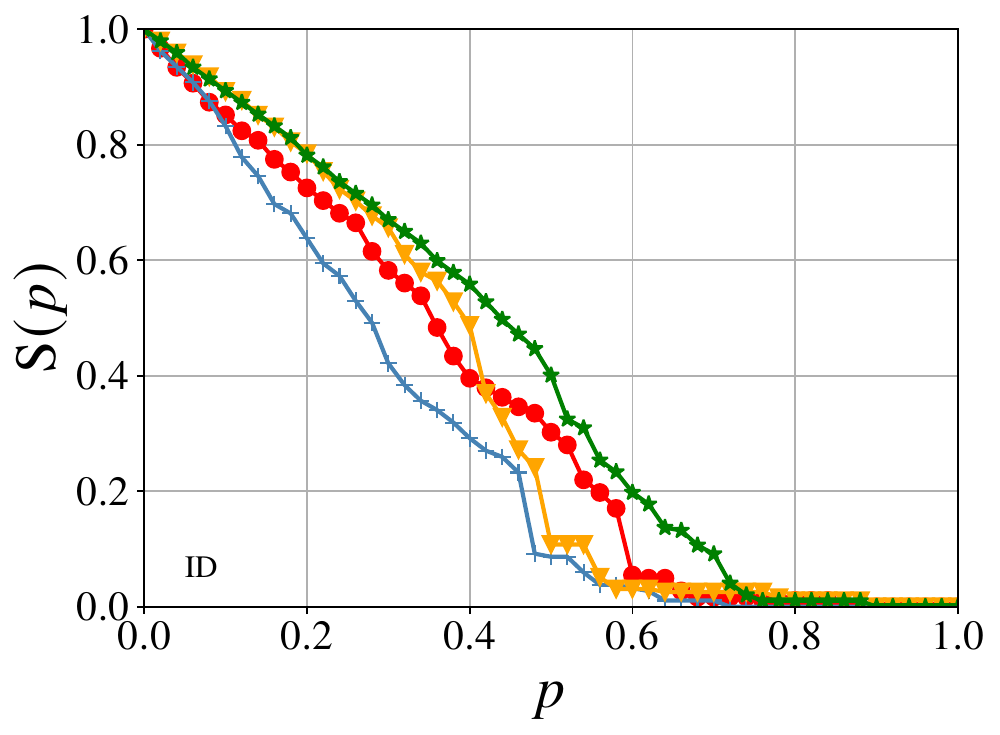}
    \includegraphics[width=0.32\linewidth]{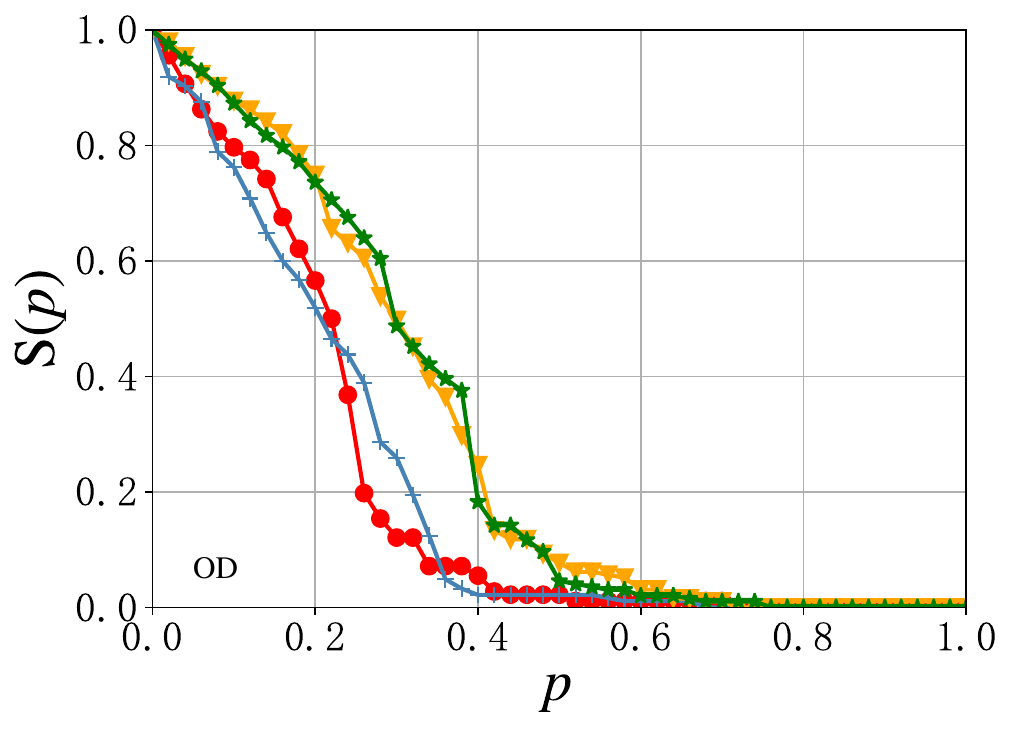}\\
    % \vspace{-3mm}
    \includegraphics[width=0.32\linewidth]{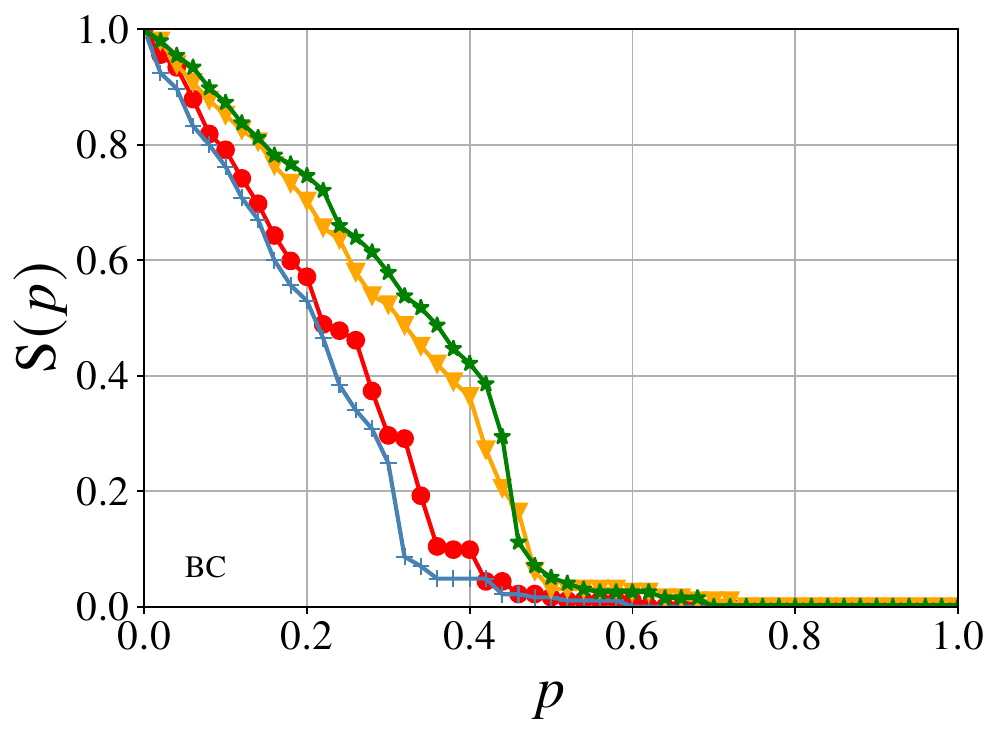}
    \includegraphics[width=0.32\linewidth]{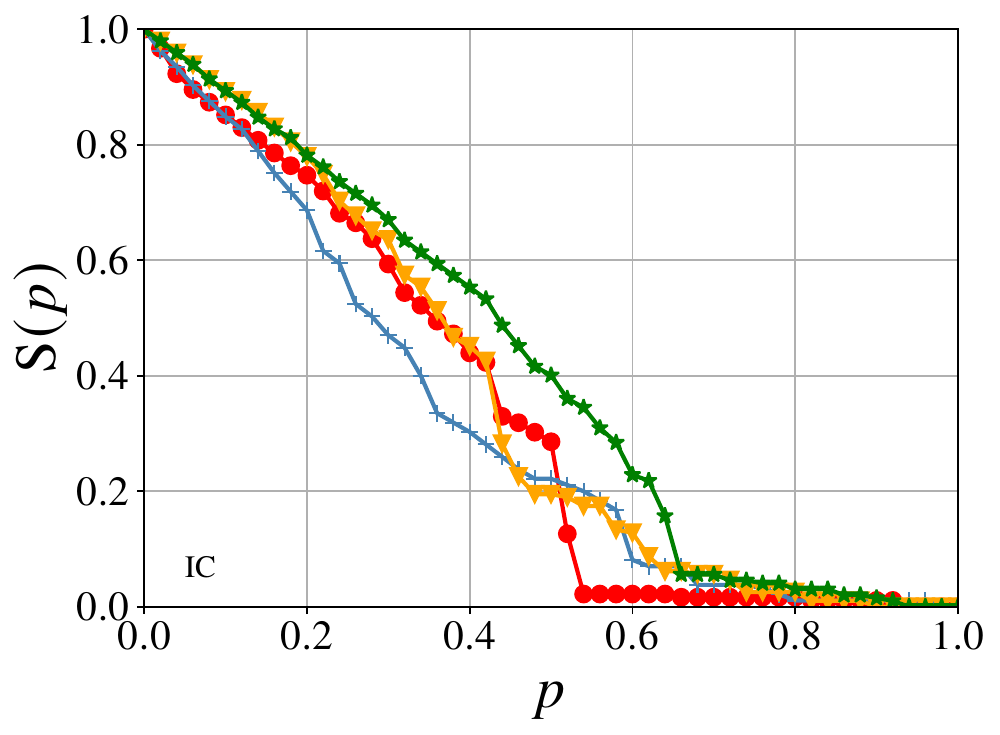}
    \includegraphics[width=0.32\linewidth]{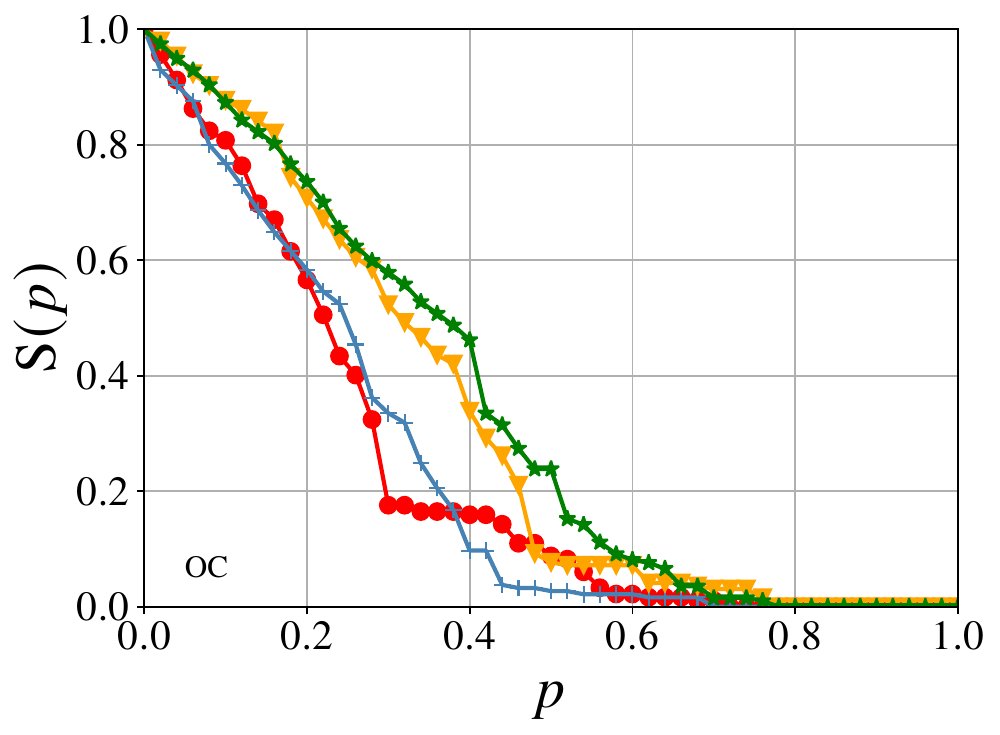}
    % \vspace{-4mm}
    \includegraphics[width=0.32\linewidth]{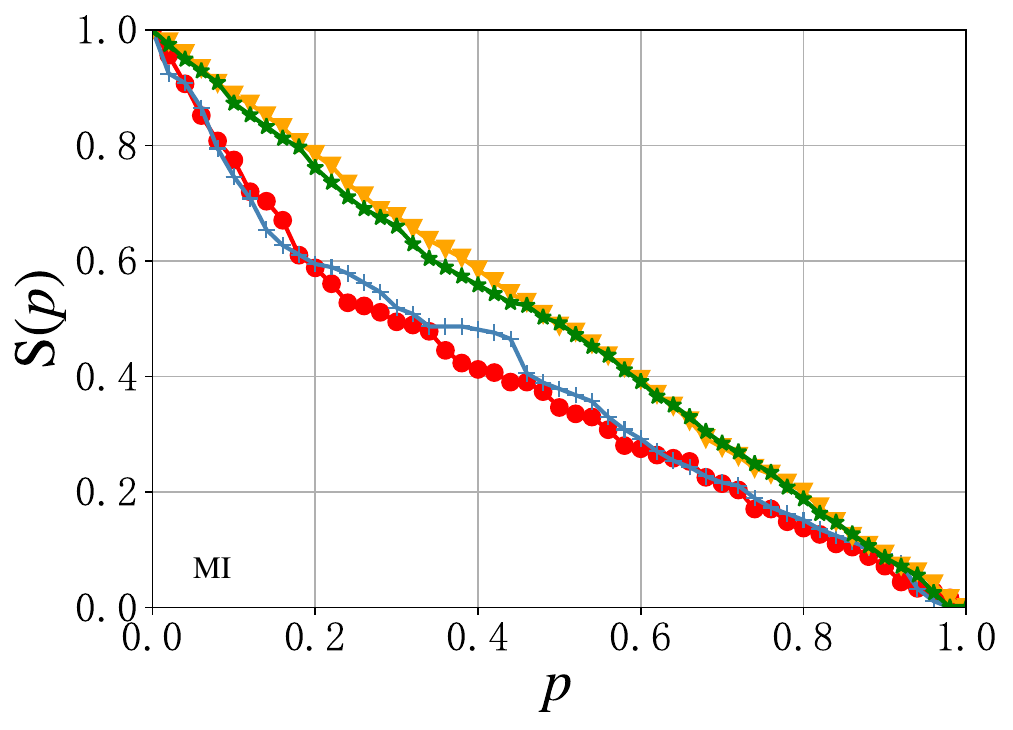}
    \includegraphics[width=0.32\linewidth]{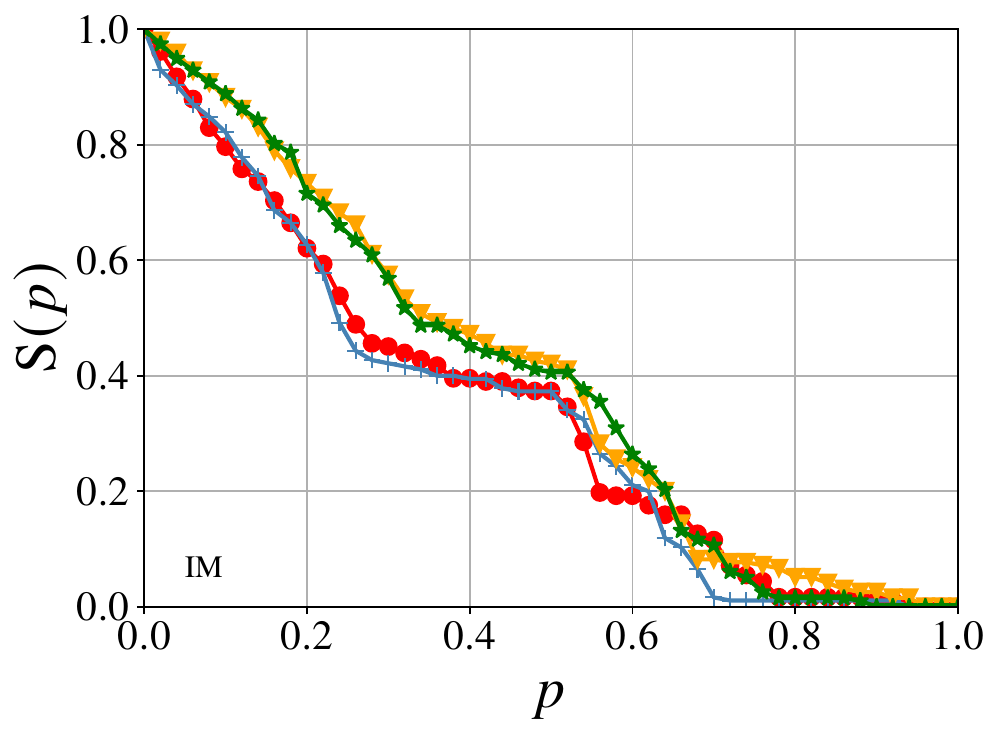}
    \includegraphics[width=0.32\linewidth]{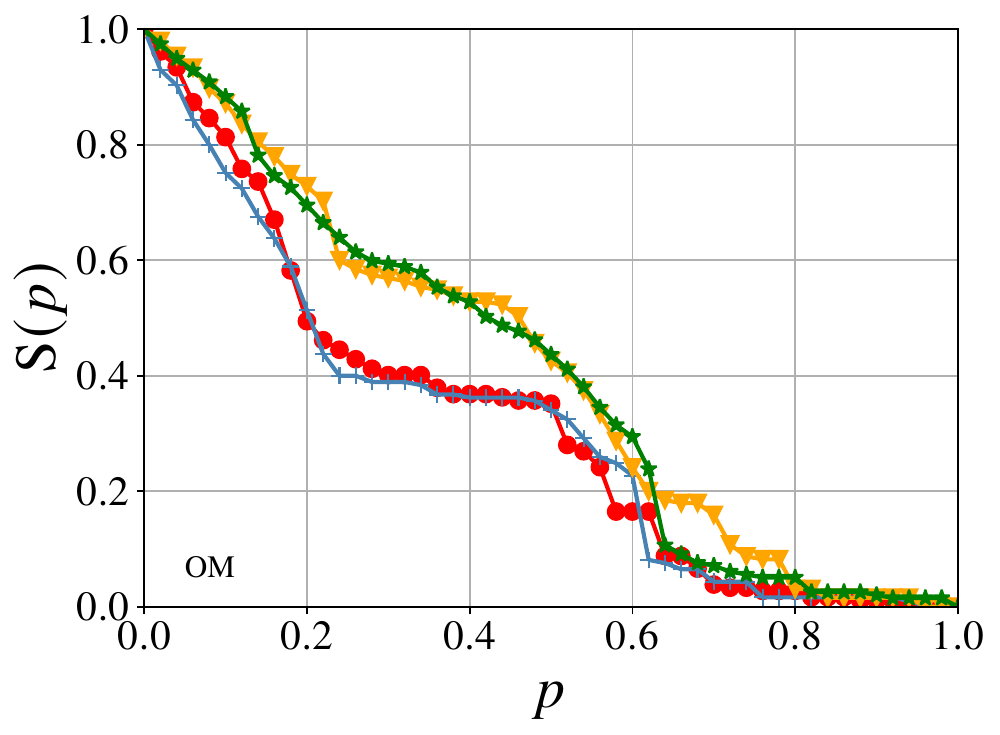}
    % \vspace{-4mm}
    \includegraphics[width=0.32\linewidth]{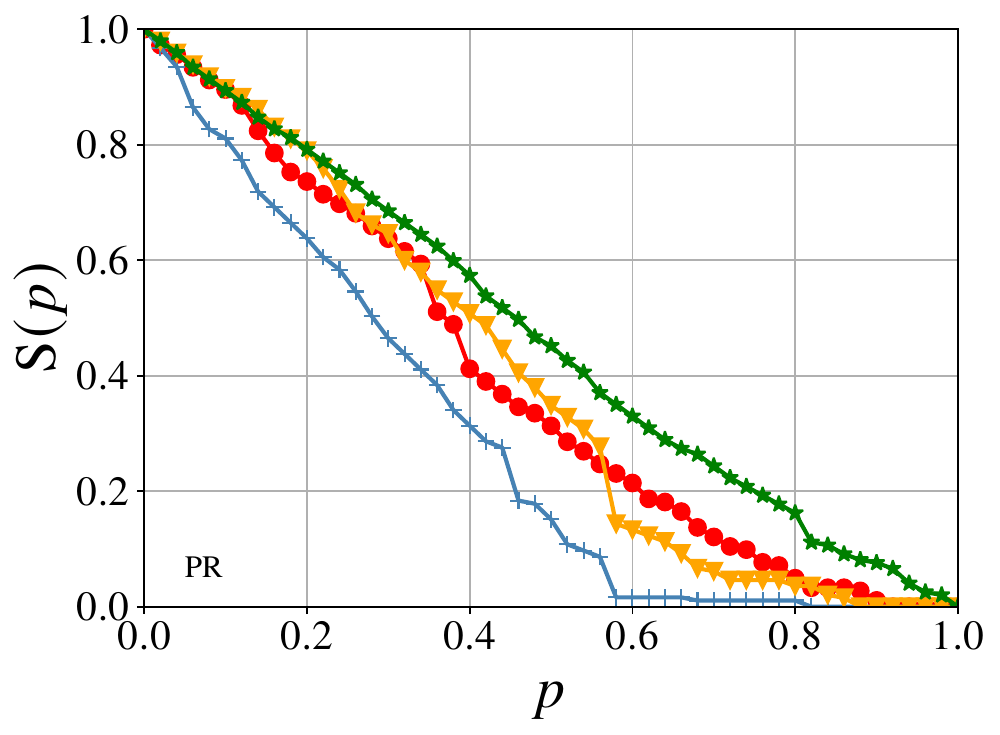}
    \includegraphics[width=0.32\linewidth]{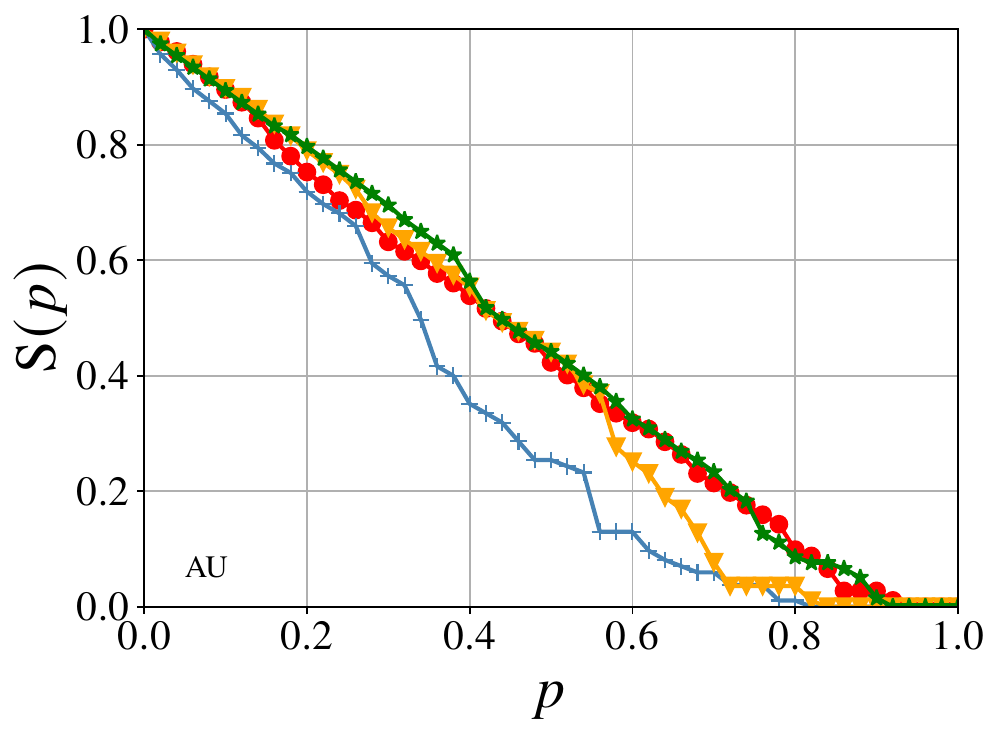}
    \includegraphics[width=0.32\linewidth]{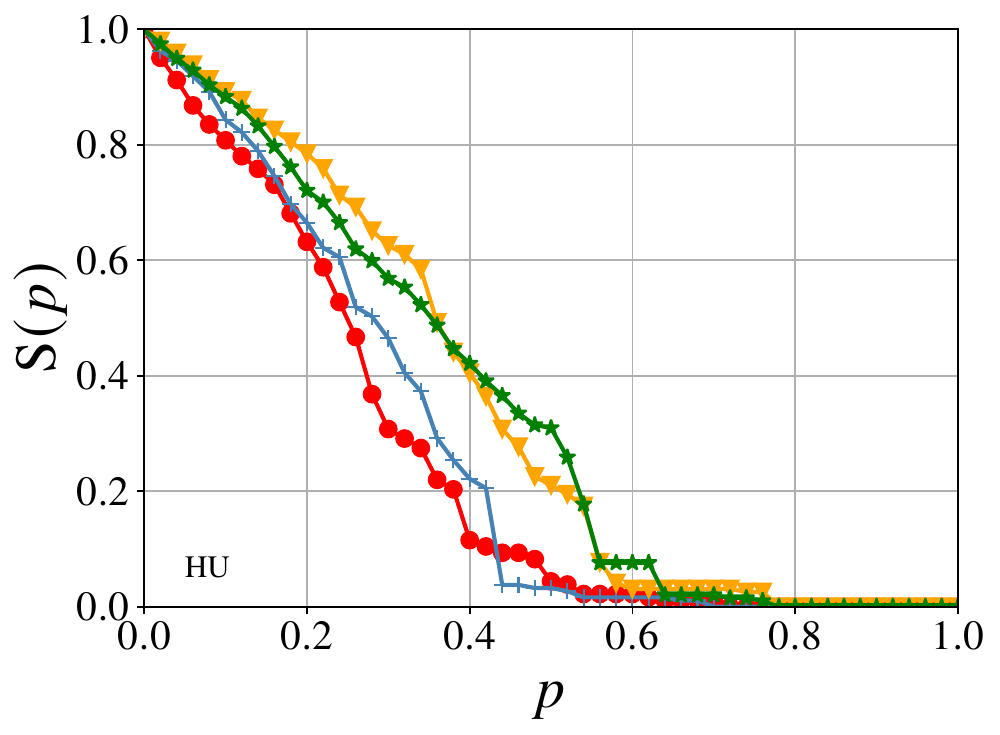}
    % \vspace{-4mm}
    \caption{Structural robustness of the four staple food supply networks under shocks to economies.}
    \label{Fig:NodeAttack_4staple}
\end{figure}

To further dissect these dynamics, we analyzed the individual networks of the four staple foods, performing strategic shocks on each. Fig.~\ref{Fig:NodeAttack_4staple} illustrates the outcomes of strategic shocks on the supply networks of four major staple crops---maize, soybeans, wheat, and rice---under shock severity of $q=100\%$. In comparison to the aggregated international food supply network, the individual networks of these staple crops exhibit significantly lower robustness. This finding underscores the stabilizing effect of aggregation, whereby the complementary nature of the four staple crops enhances the robustness of the overall food supply network.

Among the four food supply networks analyzed, the soybean network exhibits the weakest robustness, collapsing first as the broadness of strategic shocks expands. Contrary to expectations, the wheat supply network ranks second in low robustness, surpassing the maize supply network in fragility. This heightened fragility can likely be attributed to the concentrated nature of wheat production and its heavy reliance on international trade. As a result, targeted disruptions to nodes with high export significance disproportionately undermine the structural robustness of the wheat supply network. In contrast, the rice supply network demonstrates exceptional robustness. 

% Fig.~\ref{Fig:Degree distribution} presents the degree distributions of the aggregated food supply network (iFSN) and the four staple food supply networks. The iFSN degree distribution appears relatively uniform, reflecting a balanced connectivity across its nodes. In comparison, the degree distributions of the four staple foods reveal varying degrees of power-law characteristics. Among them, the soybean network displays the clearest power-law distribution, characterized by a broad range of high-frequency, low-degree nodes, signifying a pronounced reliance on a few critical nodes. Conversely, the rice network exhibits the most dispersed degree distribution, indicating a relatively even connectivity structure.

% \begin{figure}[h!]
%     \flushleft 
%     \includegraphics[width=0.32\linewidth]{All_degree_distribution.pdf}
%     \includegraphics[width=0.32\linewidth]{Corn_degree_distribution.pdf}
%     \includegraphics[width=0.32\linewidth]{Soybean_degree_distribution.pdf}\\
%     \includegraphics[width=0.32\linewidth]{Wheat_degree_distribution.pdf}
%     \includegraphics[width=0.32\linewidth]{Rice_degree_distribution.pdf}
%     \caption{The degree distribution of networks.}
%     \label{Fig:Degree distribution}
% \end{figure}

\begin{figure}[h!]
    \centering
    \includegraphics[width=0.45\linewidth]{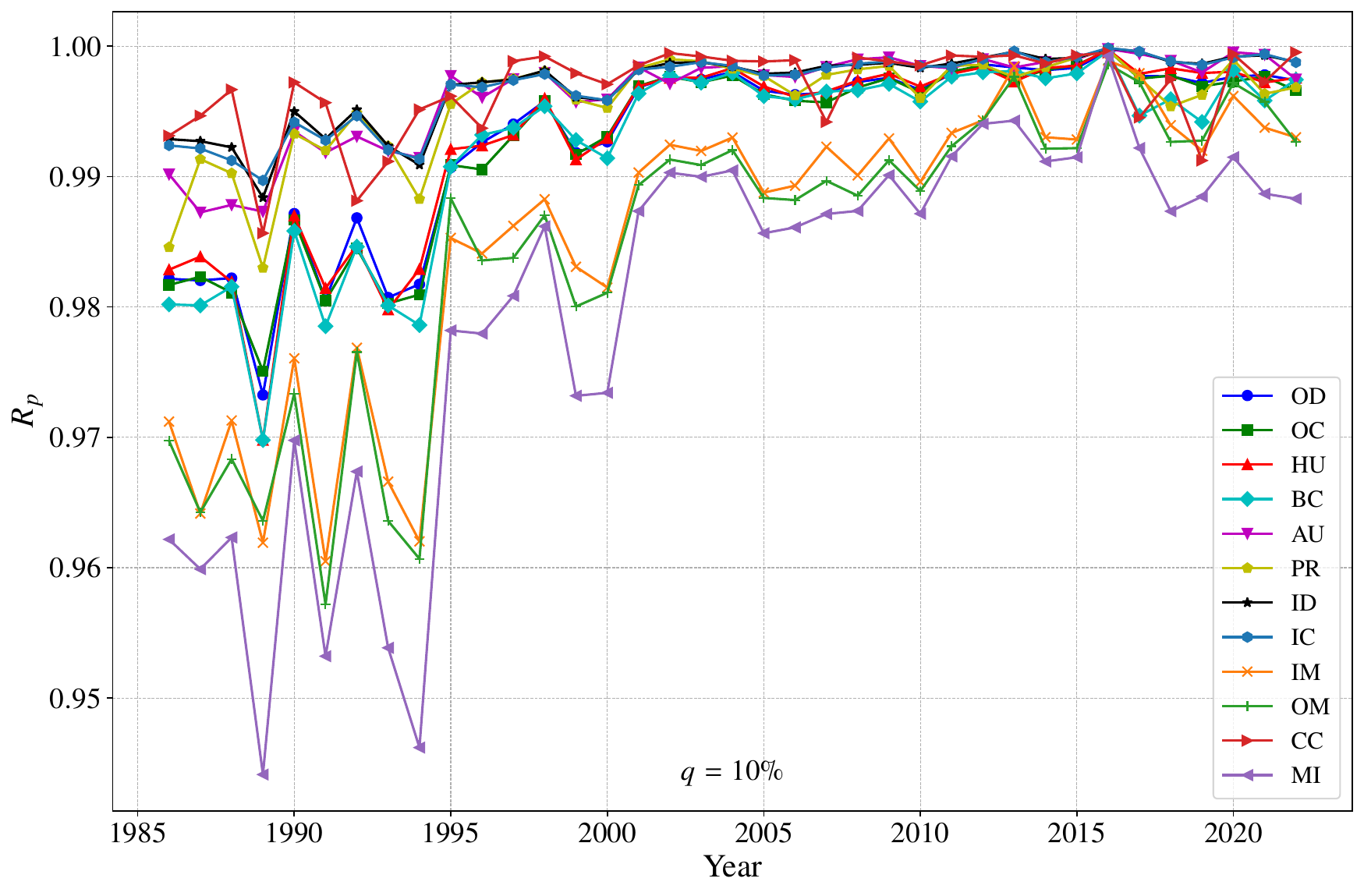}
    \includegraphics[width=0.45\linewidth]{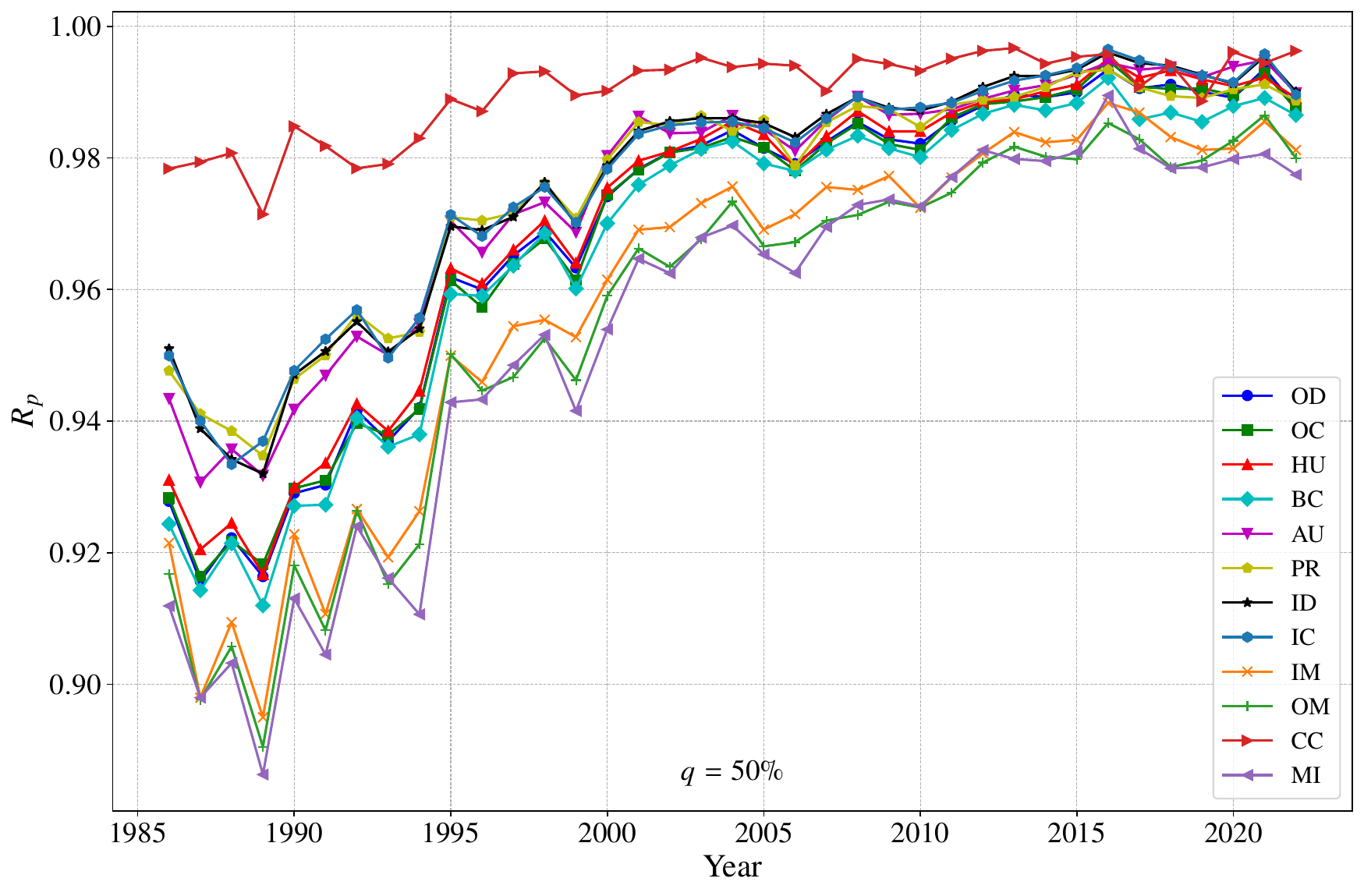}\\
    \includegraphics[width=0.45\linewidth]{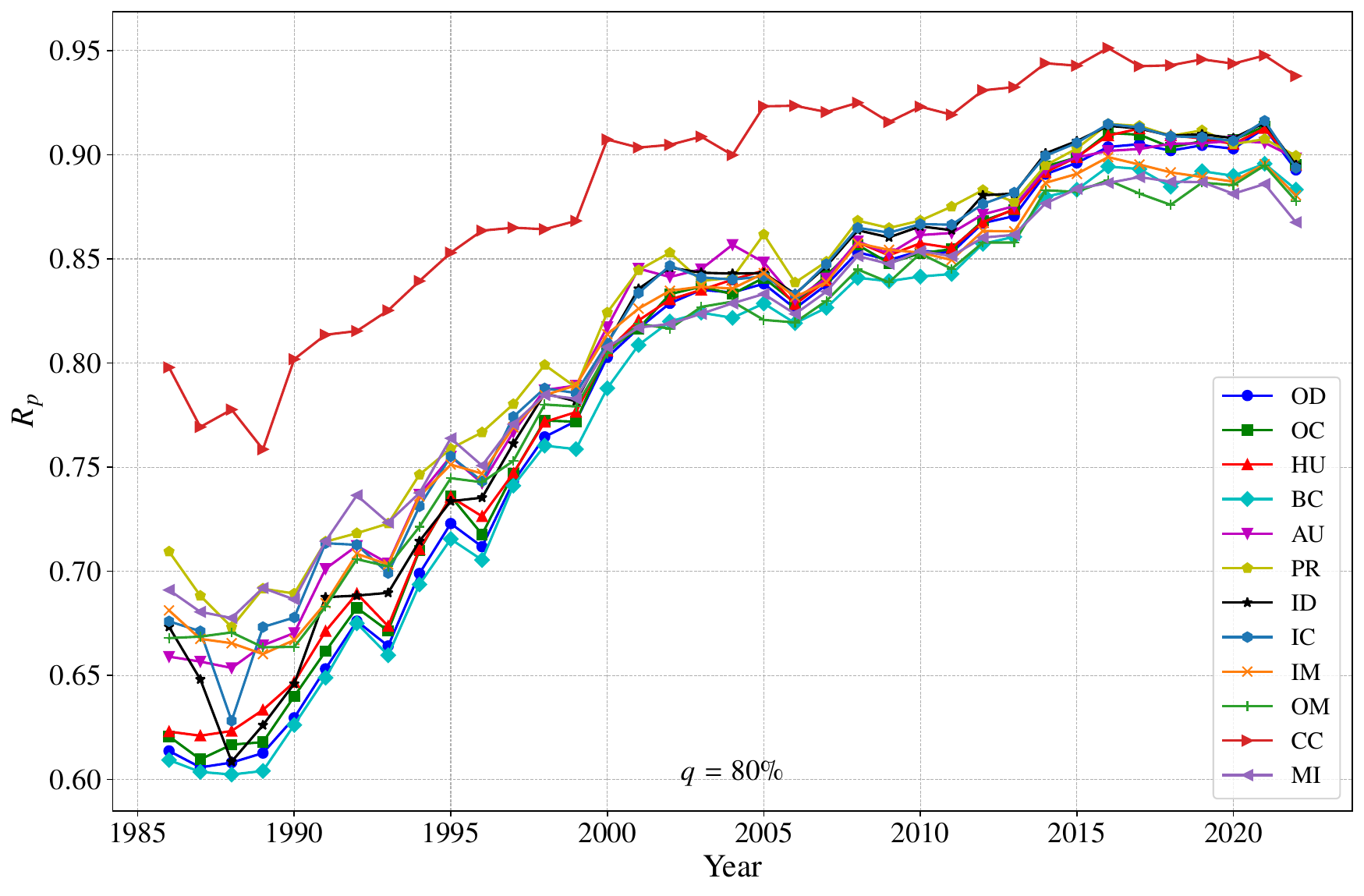}
    \includegraphics[width=0.45\linewidth]{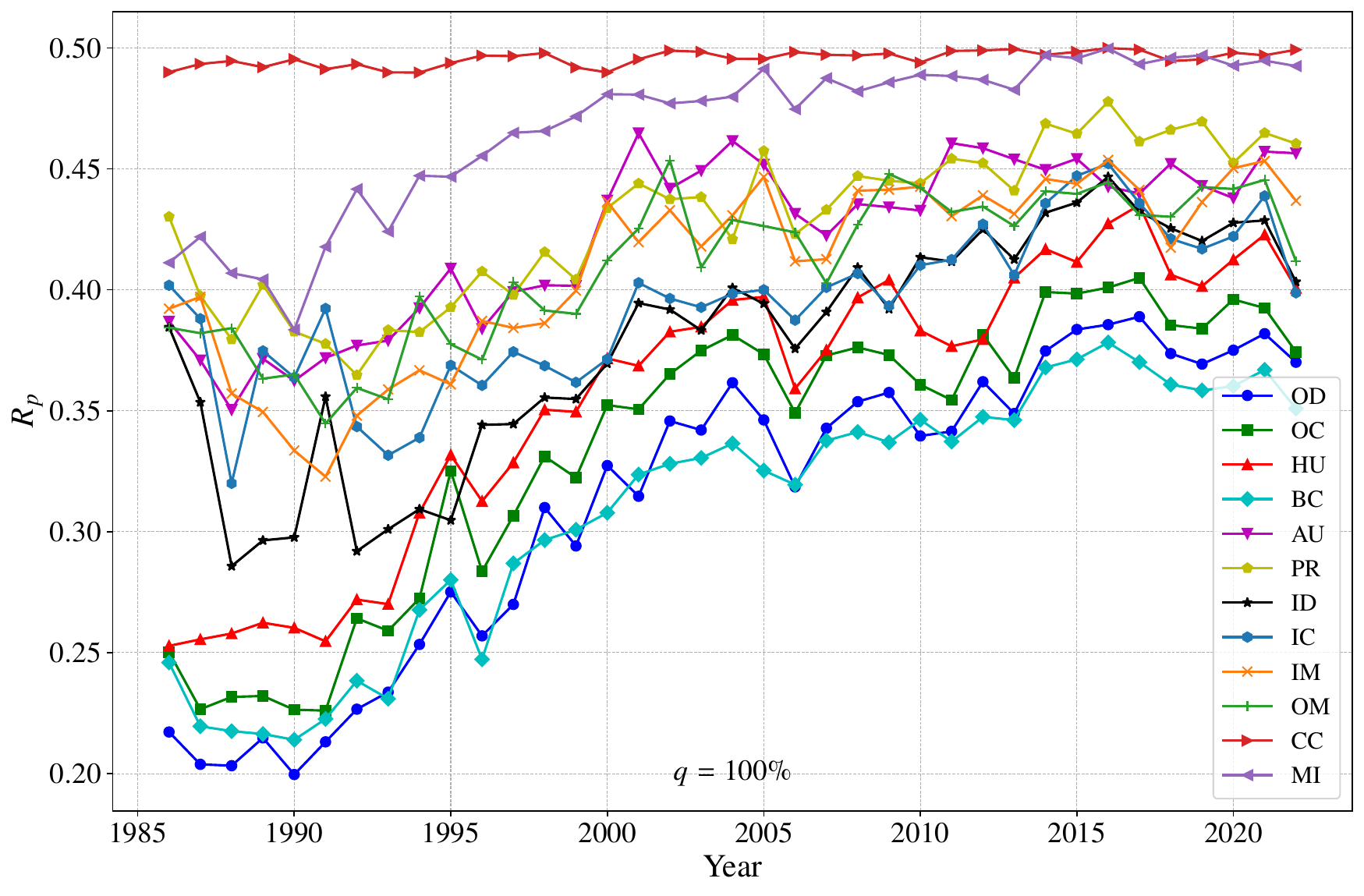}
    \caption{The robustness evolution of the iFSN from 1986 to 2022 for $q=10\%, 50\%, 80\% $ and $100\%$.}
    \label{Fig:Network_p_evolution}
\end{figure}

Fig.~\ref{Fig:Network_p_evolution} illustrates the evolution of network robustness under shock severities of $q = 10\%, 50\%, 80\%$, and $100\%$ from 1986 to 2022.
The analysis reveals several key findings. First, across all levels of shock severity, the structural robustness of the network exhibits a noticeable improvement over time within the sample interval. Second, under lower shock severities ($q = 10\%$ and $50\%$), network robustness consistently remains high, with robustness indices exceeding 0.88 throughout the period. At moderate and high shock severities ($q = 50\%$ and $80\%$), significant improvements in robustness are observed between 1990 and 2000, likely reflecting the impacts of globalization and the widespread adoption of information technology. At the highest shock severity ($q = 100\%$), network robustness remains below 0.5 throughout the entire interval, highlighting the serious fragility of the food supply network to severe disruptions.
Additionally, except at the lowest shock severity ($q = 10\%$), the network exhibits its greatest robustness when subjected to sequential shocks based on clustering coefficient rankings, indicating this method has the weakest disruptive effect. At $q = 10\%$ and $50\%$, shocks targeting nodes ranked by mutual information (MI) are the most effective, while at $q = 80\%$ and $100\%$, shocks targeting nodes ranked by betweenness centrality (BC) have the most pronounced impact.

\begin{figure}[h!]
    \centering
    \includegraphics[width=0.6\linewidth]{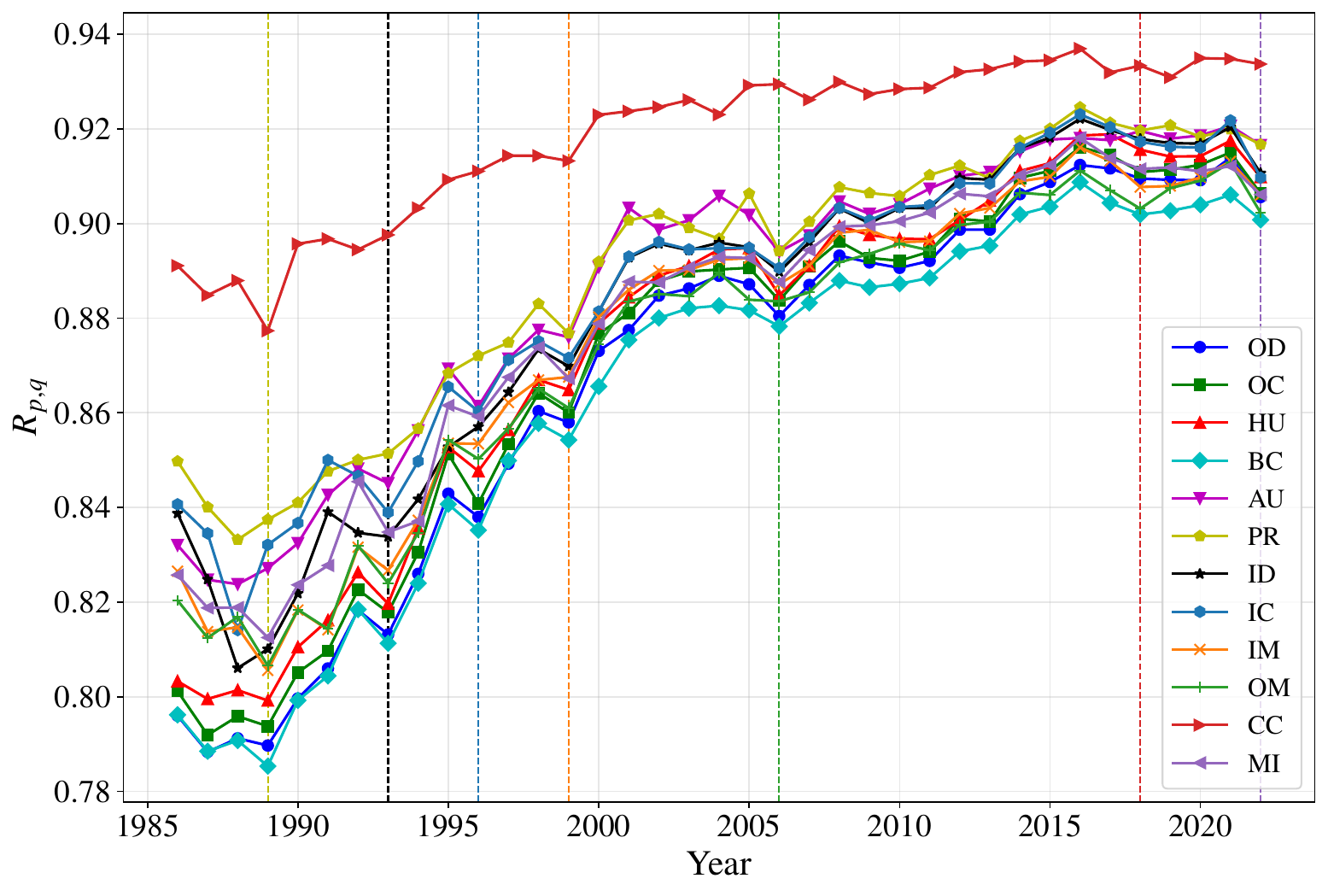}
    \caption{The robustness evolution of the iFSN measured by volume in three-dimensional coordinates, 1986-2022.}
    \label{Fig:Network_pq_evolution}
\end{figure}

In accordance with Eq.~(\ref{Eq:Robustness_volumn}), the network robustness of the international food supply network (iFSN) from 1986 to 2022 was calculated. The results are presented in Fig.~\ref{Fig:Network_pq_evolution}. As shown in the figure, robustness exhibits a steady increase within the sampled interval. This trend may be attributed to the growing number of economies integrating into the food supply network, along with the strengthening of interconnections between them, leading to an overall increase in network density. Notably, aside from the clustering coefficient, which represents the weakest node importance metric, the robustness values and trends across other indicators exhibit consistency. This alignment underscores the validity of using the volume enclosed by $p$, $q$, and $S(p,q)$ as a robust measure for assessing the structural stability of the network.

As illustrated in Fig.~\ref{Fig:Network_pq_evolution}, the robustness of the food supply network experienced significant declines in 1989, 1993, 1996, 1999, 2006, 2018, and 2022. The pronounced drop in 1989 can be attributed to geopolitical uncertainties, including turbulence preceding the end of the Cold War, compounded by the severe impacts of a strong El Ni{\~{n}}o event. In 1993, historic flooding in the Midwest United States disrupted soybean and maize production, leading to sharp reductions in export capacity. In 1996, the Uruguay Round of agricultural trade agreements began to be gradually implemented, global grain production decreased, and the inventory of major exporting countries decreased. In 1999, recurring La Ni{\~{n}}o events adversely affected agricultural yields. 
The drop in 2006 stemmed from a surge in oil prices, which prompted the United States to divert significant amounts of maize towards ethanol production, driving up agricultural prices and tightening global food supplies. In 2018, escalating trade tensions between China and the United States severely impacted U.S. exports of soybeans and maize to China, disrupting the global food trade network. Most recently, in 2022, the outbreak of the Russia–Ukraine conflict led to severe disruptions in the global food supply chain.
Taken together, these findings highlight that the robustness of the food supply network is predominantly influenced by climate variability, agricultural production, policy uncertainties, and geopolitical disruptions.

\section{Determinants of connectedness robustness}
\label{S5:EmpAnal2}

We investigate the factors influencing the structural robustness of agricultural product supply networks across three key stages: production, transportation, and trade.
At the production stage, agricultural production serves as the primary driver of the supply network, directly influencing both the scale and stability of supply. \cite{Renard-Tilman-2019-Nature} highlighted that a decline in agricultural output undermines food access. So we choose production (PROD) as an explanatory factor. We hypothesize that higher food production strengthens the structural robustness of the food supply network.
At the transportation stage, fluctuations in transportation costs significantly impact the international flow of agricultural products. High transportation costs can lead to supply disruptions. \cite{Mian-Huse-Borruel-Doumeizel-2020-CerealFoodsWorld} emphasized that the entire food supply chain relies on the reliable functioning of the transportation system, leading us to select the Baltic Dry Index (BDI) as a key explanatory variable. We posit that increased transportation costs correspond to greater instability in food supply.
At the trade stage, food prices play a pivotal role in shaping trade dynamics. \cite{Lagi-BarYam-Bertrand-BarYam-2015-ProcNatlAcadSciUSA} demonstrated that rising basic food prices have profound implications for food security, particularly among vulnerable populations. Sharp fluctuations in food prices can introduce market uncertainty, further destabilizing the supply network. Consequently, we include the food price index (FPI) as an explanatory factor.
In addition to these factors that influence each stage directly, certain variables affect multiple stages of the supply chain. Extreme climate events, for example, can disrupt both production and transportation, thereby threatening food security. Previous studies have documented the dual impact of global natural disasters on both agricultural production and transportation \citep{Parven-Pal-Witayangkurn-Pramanik-Nagai-Miyazaki-Wuthisakkaroon-2022-IntJDisasterRiskReduct, Gossling-Neger-Steiger-Bell-2023-NatHazards}. As such, we incorporate global natural disasters (GND) as an explanatory variable.
Oil prices have far-reaching effects on production costs, transportation expenses, and trade prices \citep{Wei-Qiu-An-Zhang-Li-Guo-2024-IntRevEconFinanc,Hung-2021-ResourPolicy}. We therefore introduce the oil price uncertainty index (OPU) as an additional explanatory variable.
Furthermore, geopolitical risks influence various stages of the agricultural product supply chain. A growing body of literature has explored the impact of geopolitical risks on global supply chains \citep{Ren-Mu-Wang-Yue-Li-Du-Zhao-Lim-2024-ComputIndEng, Xiong-Wu-Yeung-2024-IntJProdRes}. Specifically, in the context of agricultural supply, \cite{Li-Wang-Kharrazi-Fath-Liu-Liu-Xiao-Lai-2024-FoodSecur} found that geopolitical factors negatively affect the resilience of agricultural trade. Accordingly, we incorporate the geopolitical risk index (GPR) as an explanatory variable.

We therefore construct a measurement model to assess the influence of these factors:
\begin{equation}
 R = f(\text{BDI}, \text{GPR}, \text{PROD}, \text{GND}, \text{FPI}, \text{OPU}), 
\label{Eq:Regression}
\end{equation}
which is the basic model in our analysis. Upon further investigation, we found that agricultural production has a significant impact on the robustness of the food supply network. Consequently, we aim to explore which crops' production and which economies' production levels most influence network robustness. As a result, we replace total production with the production of the four main staple crops, as well as the production of these crops from the top five economies listed in Table~\ref{Table: Top 10 freq}.

We employ multiple models to explore the relationship between the robustness of the iFSN and the aforementioned factors.
First, we apply multiple OLS regression for an initial linear analysis. However, to address potential multicollinearity and overfitting, we introduce stepwise regression, which iteratively selects variables to optimize the model. Despite its advantages, multicollinearity persists. To mitigate this, we use ridge regression, which adds a regularization term to the loss function, improving stability by reducing multicollinearity and overfitting. 
While the first three models assume linear relationships, the effects of these factors on network robustness may be nonlinear. To account for this, we introduce the random forest model, a machine learning technique based on ensemble learning with decision trees. By constructing multiple decision trees and aggregating their predictions, this model improves both accuracy and robustness.
% The results are presented in Tables~\ref{tab:Regression_all}--\ref{tab:Regression_5economies}.

\subsection{Basic model}

Table~\ref{tab:Regression_all} presents the correlation coefficients, multivariate regression results, and random forest feature importance rankings for key potential drivers of network robustness in the baseline model. The correlation coefficients between explanatory variables and network robustness indicate that total grain production exhibits the strongest correlation, followed by the food price index and the oil price uncertainty index. This suggests that these three factors may play a significant role in shaping network robustness, warranting further discussion in conjunction with regression results. 
Prior to regression analysis, all variables were standardized to ensure comparability of coefficients. Both OLS and stepwise regression identified the Baltic Dry Index (BDI), total food production (PROD), global natural disasters (GND), and the food price index (FPI) as significant determinants of food supply network robustness. Stepwise regression, employed to mitigate multicollinearity, retained only these four key drivers. Among them, PROD exhibited the strongest positive effect, while FPI had a negative impact---despite its initial positive correlation, likely due to controlling for grain production. BDI and GND showed modest positive associations, with BDI primarily reflecting shipping market dynamics (rather than transport costs) and GND’s influence tempered by its geographically dispersed nature.
These findings were further supported by ridge regression. Additionally, random forest feature importance analysis confirmed PROD as the strongest contributing factor, with substantially higher importance scores than other variables, underscoring food production’s dominant role in network robustness.

Taken together, these findings establish total food production as the dominant factor shaping food supply network stability. Building upon the baseline model, we further disaggregate this factor to examine the influence of specific crop types and food production across different economies, thereby gaining deeper insights into their respective contributions to network robustness.

\begin{table}[htbp]
   \centering
   \caption{Regression results of influencing factors to network robustness. }
   \newcolumntype{d}[1]{D{.}{.}{#1}} % 定义小数点对齐列
   \begin{tabular*}{\textwidth}{@{\extracolsep{\fill}}c d{2.5} *{2}{d{2.10}} d{2.5} d{1.5} d{1.5} @{}}
   \toprule
   & \multicolumn{1}{c}{\textbf{Correlation}}
   & \multicolumn{1}{c}{\textbf{OLS}} 
   & \multicolumn{1}{c}{\textbf{Stepwise}} 
   & \multicolumn{1}{c}{\textbf{Ridge}} 
   & \multicolumn{1}{c}{\textbf{RF importance}}
\\
   \midrule
   BDI        & 0.2149^{**} & 0.0087\,(0.003)^{**}  & 0.0085\,(0.003)^{***}  & 0.0082  &0.0052 \\
   GPR        & 0.0206^{*} & 0.0004\,(0.003)  &                & -0.0005 &   0.0428 \\
   GND        & -0.0084^{*} & 0.0057\,(0.003)^{*}  & 0.0059\,(0.003)^{**}  & 0.0048  & 0.0099 \\
   FPI        & 0.6517^{***} & -0.0099\,(0.005)^{*} & -0.0100\,(0.005)^{**} & -0.0077  &  0.0233 \\
   OPU        & 0.3653^{***} & -0.0006\,(0.004) &               & -0.00004  & 0.0383 \\
   PROD       & 0.8817^{***} & 0.0437\,(0.005)^{***} & 0.0436\,(0.005)^{***} & 0.0403  & 0.8804 \\
   \bottomrule
   \end{tabular*}
    \begin{tablenotes}   
        \footnotesize        
        \item Note: Estimates for the intercept are not reported, and standard errors are shown in parentheses. * indicates significance at the 10\% level; ** at the 5\% level; *** at the 1\% level.
      \end{tablenotes}  
   \label{tab:Regression_all}
\end{table}

\subsection{The impact of staple food type}

Table~\ref{tab:Regression_4staple} presents the correlation coefficients between crop productions and network robustness, followed by the regression results using various crop productions as explanatory variables, along with the feature importance ranking from random forest analysis.
The correlation coefficients show that the productions of the four staple crops are highly correlated with the robustness of the food supply network, all exceeding 0.8. 
However, regression results indicate that among the four crop productions, only rice production has a statistically significant positive coefficient, suggesting that rice production plays a crucial and positive role in the stability of the food supply network. The other factors align with the baseline model. In the random forest importance ranking, soybean production emerges as the most influential factor, followed by rice production, indicating that the impact of soybean production on the robustness of the food supply network is nonlinear. This may be attributed to the fact that soybeans are not only consumed as food but also widely used as livestock feed and for industrial purposes. Moreover, while maize, soybeans, and rice exhibit relatively similar levels of contribution, wheat's importance is considerably lower than that of the other three crops. This may be due to the more decentralized nature of global wheat production, along with its greater substitutability.

\begin{table}[htbp]
\centering
\caption{Regression results of network robustness with different staple food production as explanatory variables.}
\label{tab:Regression_4staple}
\newcolumntype{d}[1]{D{.}{.}{#1}} % 定义小数点对齐列
\begin{tabular*}{\textwidth}{@{\extracolsep{\fill}}c d{1.5} *{2}{d{2.10}} d{2.5} d{1.5}  @{}}
% \begin{tabular}{c *{2}{d{2.10}} d{2.5} d{1.5} @{}}
\toprule
 & \multicolumn{1}{c}{\textbf{Correlation}}
 & \multicolumn{1}{c}{\textbf{OLS}} 
 & \multicolumn{1}{c}{\textbf{Stepwise}} 
 & \multicolumn{1}{c}{\textbf{Ridge}} 
 & \multicolumn{1}{c}{\textbf{RF importance}}
\\
\midrule
BDI                & 0.2149^{**} & 0.0052\,(0.003)^{*}  & 0.0065\,(0.002)^{***}  & 0.0067  & 0.0032 \\
GPR                & 0.0206^{*} & 0.0030\,(0.002)  &                & 0.00004 & 0.0324 \\
GND                & -0.0084^{*} & 0.0053\,(0.002)^{**}  & 0.0062\,(0.002)^{***}  & 0.0054  & 0.0063 \\
FPI                & 0.6517^{***} & -0.0142\,(0.004)^{**} & -0.0147\,(0.003)^{***} & -0.0113  & 0.0137 \\
OPU                & 0.3653^{***} & 0.0005\,(0.003) &               & 0.0005  & 0.0207 \\
Maize              & 0.8657^{***} & -0.0137\,(0.020) &                & -0.0025  & 0.2259 \\
Soybeans           & 0.8924^{***} & 0.0010\,(0.019)  &                & 0.0071  & 0.3756 \\
Rice               & 0.9171^{***} & 0.0616\,(0.011)^{***} & 0.0489\,(0.003)^{***} & 0.0309  & 0.3016 \\
Wheat              & 0.8378^{***} & -0.0009\,(0.009) &                & 0.0098  & 0.0207 \\
\bottomrule
\end{tabular*}
    \begin{tablenotes}   
        \footnotesize        
        \item Note: Estimates for the intercept are not reported, and standard errors are shown in parentheses. * indicates significance at the 10\% level; ** at the 5\% level; *** at the 1\% level.
      \end{tablenotes}  
\end{table}

% \newpage
\subsection{The impact of economies' production}

To examine the impact of the production of different economies on the robustness of the global food supply network, we selected the top five economies' food production from Table~\ref{Table: Top 10 freq} as explanatory variables in the model. The correlation coefficients, followed by the regression results and the feature importance rankings from the random forest analysis, are presented in Table~\ref{tab:Regression_5economies}.

Analysis reveals US, Indian, and Chinese food production show the strongest correlation with network robustness. Regression results indicate all five major economies except Canada significantly impact the supply network, as Canada's production volume and diversity are comparatively limited. Notably, when including these economies' production data, previously significant factors like the Baltic Dry Index become insignificant, suggesting dominant producers' stable logistics systems outweigh broader market indicators.
India and the US demonstrate strongly positive effects on robustness, while China shows a negative impact---likely due to its domestic-focused food policies and reliance on imports, decoupling its production from global demand dynamics. Random forest analysis confirms these findings, highlighting India's critical role as a top rice exporter (particularly influencing Asia and Africa) and the US's dominance in wheat and maize markets, where production fluctuations directly affect global price stability and supply chain robustness.

\begin{table}[htbp]
\centering
\caption{Regression results of network robustness with different economies' staple food production as explanatory variables.}
\label{tab:Regression_5economi}
\newcolumntype{d}[1]{D{.}{.}{#1}} % 定义小数点对齐列
\begin{tabular*}{\textwidth}{@{\extracolsep{\fill}}c d{1.5} *{2}{d{2.10}} d{2.5} d{1.5}  @{}}
% \begin{tabular}{c *{2}{d{2.10}} d{2.5} d{1.5} @{}}
\toprule
 & \multicolumn{1}{c}{\textbf{Correlation}}
 & \multicolumn{1}{c}{\textbf{OLS}} 
 & \multicolumn{1}{c}{\textbf{Stepwise}} 
 & \multicolumn{1}{c}{\textbf{Ridge}} 
 & \multicolumn{1}{c}{\textbf{RF importance}}
 \\
\midrule
BDI                & 0.2149^{**} & 0.0011\,(0.003)  &                 & 0.0026     & 0.0034  \\
GPR                & 0.0206^{*} & 0.0012\,(0.003)  &                 & 0.0014     & 0.0517  \\
GND                & -0.0084^{*} & 0.0053\,(0.002)^{**}  & 0.0063\,(0.002)^{***}  & 0.0029     & 0.0042  \\
FPI                & 0.6517^{***} & -0.0089\,(0.004)^{**} & -0.0088\,(0.004)^{**} & -0.0069    & 0.0193  \\
OPU                & 0.3653^{***} & 0.0028\,(0.003)  &                 & 0.0032     & 0.0501  \\
CAN                & 0.6261^{***} & -0.0032\,(0.005) &                 & -0.0067    & 0.0057  \\
CHN                & 0.8137^{***} & -0.0139\,(0.010) & -0.0205\,(0.007)^{***} & 0.0078 & 0.0200  \\
FRA                & 0.5776^{***} & 0.0075\,(0.002)^{***} & 0.0076\,(0.002)^{***}  & 0.0083     & 0.0359  \\
IND                & 0.8795^{***} & 0.0378\,(0.009)^{***} & 0.0414\,(0.007)^{***} & 0.0189     & 0.7198  \\
USA                & 0.8910^{***} & 0.0196\,(0.005)^{***} & 0.0196\,(0.005)^{***} & 0.0177     & 0.0899  \\
\bottomrule
\end{tabular*}
 \begin{tablenotes}   
        \footnotesize        
        \item Note: Estimates for the intercept are not reported, and standard errors are shown in parentheses. * indicates significance at the 10\% level; ** at the 5\% level; *** at the 1\% level.
      \end{tablenotes}  
\label{tab:Regression_5economies}
\end{table}

\section{Conclusions}
\label{S6:Conclude}

The structural robustness of the international food supply network is crucial for ensuring food security across nations. In this study, we construct an aggregated food supply network by weighting four major staple crops based on caloric contribution and analyze its structural robustness. We first assess the influence of different economies within the network using various centrality metrics. Then, we conduct strategic shocks on both the aggregated network and individual staple crop networks, ranked by economic importance, to evaluate the structural robustness of the food supply network under disruption. Furthermore, we employ regression analysis and random forest models to identify key factors affecting network robustness. In this section, we synthesize our findings, discuss their theoretical and practical implications, and outline potential directions for future research.

\subsection{Main findings}

As economies place growing importance on food security, they have increasingly diversified their food sources, leading to a steady rise in the density of the global food supply network from 1986 to 2022. Our study first examines the 2022 food supply network, ranking economies based on their significance within the iFSN. We identify the Netherlands as the primary food consumer, while India and the United States emerge as the dominant suppliers. Island nations and African countries exhibit strong regional dependencies. Overall, the most critical economies in the food supply network are the United States, India, China, and France, highlighting that suppliers play a more pivotal role in network stability than consumers.

Simulated network shocks based on different importance rankings indicate that the international food supply network demonstrates strong robustness. Only when both the severity and broadness of shocks are significantly high does the iFSN structure experience severe disruption. Furthermore, the decline in network robustness under shock exhibits two distinct characteristics. First, the degradation follows an initial slow decline before a sharp drop, attributed to the network’s high overall connectivity and redundancy. Second, when shock severity reaches 100\%, the robustness decline initially follows a linear pattern, suggesting that the top-ranked economies contribute relatively evenly to network robustness.
Further analyses of the four staple crop supply networks reveal that these individual networks are more vulnerable than the aggregated iFSN. Among them, the soybean supply network exhibits the weakest robustness, followed by wheat, maize, and rice. 

This study evaluates the robustness of networks in both two-dimensional and three-dimensional frameworks under simulated shocks. Analyzing the evolution of two-dimensional robustness reveals that the network maintains a high level of stability under low-severity shocks. At 50\% and 80\% shock severity, the network exhibits a marked improvement in robustness between 1990 and 2000. However, extreme shocks significantly undermine the robustness of the global food supply network.
Extending the analysis to three-dimensional robustness, we find that simulated shocks targeting different metrics yield highly similar trends and numerical values, indicating that this metric provides a reliable measure of the supply network’s robustness. Moreover, robustness exhibits an overall upward trend throughout the sample period. A closer examination of seven notable declines in network robustness suggests that key contributing factors include climate variability, crop yields, geopolitical events, fluctuations in food and oil prices, and transportation disruptions. This motivates a further investigation into the underlying determinants of network robustness.

Using regression analysis and random forest models, we identify food production as the most critical positive factor influencing the robustness of the global food supply network. Disaggregating food production by crop type and economic region, we find that rice and soybean production, as well as grain production in India and the United States, play a dominant role in shaping network robustness. Notably, India is a major rice exporter, aligning with our finding that the rice supply network is the most vulnerable among the four staple grains. Additionally, food prices exert a consistently negative and significant impact on network robustness across all models.

\subsection{Implications}

The implications of this study encompass both theoretical and practical dimensions. 
Theoretically, the study contributes in two key aspects.
First, by employing strategic shocks on critical nodes, this research reveals the fragilities of the global food supply network under specific disruptions. Beyond expanding the application of network shock simulations in the context of food security, it provides deeper insights into the structural robustness of food supply systems against external shocks. These findings advance theoretical frameworks on network robustness, facilitating the development of a more comprehensive methodology for assessing supply network stability. Moreover, they offer valuable theoretical references for the study of other economic networks.
Second, the study uncovers significant differences in the robustness of supply networks across different staple crops and analyzes the contributions of various staple foods and economies to the global food supply network. It further highlights the heterogeneity of food supply chains, demonstrating that variations in production models, trade structures, and the roles of key supplier nations lead to distinct network dynamics. This provides a theoretical foundation for future research on the security of different food supply chains and offers a novel methodological perspective for analyzing the roles of economies in the broader global economic network.

The practical implications of this study lie in enhancing the structural robustness of the global food supply network, which can be summarized in three key aspects.
First, governments and international organizations should foster international cooperation and promote supply chain diversification to enhance food security. This study identifies the most influential economies within the global food supply network and, through simulated shocks, demonstrates that external shocks to these key players can significantly disrupt the network. Meanwhile, within the sampled interval, the strengthening interconnections among economies in the food supply network enhanced robustness. As a result, governments and international organizations should strengthen trade cooperation with these critical countries while promoting supply diversification to enhance food security. 
Second, policymakers should develop targeted risk management strategies to address vulnerabilities in the food supply network. The study reveals the robustness characteristics of the global food supply network under extreme disruptions and identifies weaknesses in key countries and specific staple crop supply chains (such as rice). These findings provide essential guidance for international organizations and national governments in formulating food security contingency plans.
Third, governments should improve food production systems, refine storage policies, and implement price stabilization mechanisms to strengthen supply network robustness. The study highlights that food production, particularly the output of rice and soybeans, is a crucial determinant of supply network stability. Therefore, economies should refine agricultural production and storage policies to ensure food security. Additionally, the findings indicate that food prices have a significant negative impact on network stability, suggesting that fluctuations in global market prices may exacerbate uncertainties in food supply. To mitigate such risks, governments can employ fiscal subsidies, establish international food reserves, and implement other stabilizing measures to reinforce the resilience of food supply chains.

\subsection{Future research directions}

Building on the findings of this study, future research could explore several key directions. First, while this study constructs the global food supply network primarily based on trade values between economies, incorporating additional factors such as production and stock levels could lead to a more comprehensive representation of the network. Second, in developing the shock model, we account for both the broadness and severity of disruptions. However, limiting the analysis to only two parameters may not fully capture the complexities of real-world supply chain shocks. Future work could introduce additional parameters to refine the shock model, enabling a more realistic simulation of disruptions and the subsequent responses of economies. This would provide deeper insights into the resilience of the global food supply network under different stress scenarios. Lastly, integrating dynamic evolutionary models could offer a valuable avenue for assessing the network’s long-term adaptability and recovery capacity in the face of sustained shocks. Such advancements would help inform more effective policy interventions aimed at strengthening global food security.

\section*{Acknowledgment}

This work was partly supported by the National Natural Science Foundation of China (72171083) and the Fundamental Research Funds for the Central Universities.

% \section*{Data availability}

% The data set in this paper are sourced from the Wind database (https://www.wind.com.cn) and the website of the International Grains Council (https://www.igc.int).

%\newpage
%
%\bibliography{E:/papers/Auxiliary/Bibliography}
%\bibliography{E:/Auxiliary/Bibliography}
%\bibliography{Bib1, Bib2, BibRobustNet, BibRCE, BibITN}

%\bibliographystyle{plain}

% \end{CJK*}
\end{document}